\documentclass[aps,pra,reprint]{revtex4-2}
\usepackage{amsfonts} 
\usepackage{xcolor}
\usepackage{graphicx}
\usepackage{dcolumn}
\usepackage{braket}
\usepackage{url}
\usepackage{bm}

\begin{document}

\preprint{APS/123-QED}

\title{Qubits on programmable geometries with a trapped-ion quantum processor}

\author{Qiming Wu}
\email{qiming.wu@nyu.edu}
\affiliation{Department of Physics, University of California, Berkeley, CA, USA}
\author{Yue Shi}
\affiliation{Department of Physics, Princeton University, Princeton, NJ, USA}
\author{Jiehang Zhang}
\email{jzhang2022@ustc.edu.cn}
\affiliation{1. School of Physical Sciences, University of Science and Technology of China, Hefei 230026, China
}
\affiliation{2. Shanghai Research Center for Quantum Science and CAS Center for Excellence in Quantum Information and Quantum Physics, University of Science and Technology of China, Shanghai 201315, China}
\affiliation{3. Hefei National Laboratory, University of Science and Technology of China, Hefei 230088, China}



\date{\today}

\begin{abstract}


Geometry and dimensionality have played crucial roles in our understanding of the fundamental laws of nature, with examples ranging from curved space-time in general relativity to modern theories of quantum gravity. In quantum many-body systems, the entanglement structure can change if the constituents are connected differently, leading to altered bounds for correlation growth and difficulties for classical computers to simulate large systems. While a universal quantum computer can perform digital simulations, an analog-digital hybrid quantum processor offers advantages such as parallelism. Here, we engineer a class of high-dimensional Ising interactions using a linear one-dimensional (1D) ion chain with up to 8 qubits through stroboscopic sequences of commuting Hamiltonians. 
In addition, we extend this method to non-commuting circuits and demonstrate the quantum XY and Heisenberg models using Floquet periodic drives with tunable symmetries. The realization of higher dimensional spin models offers new opportunities ranging from studying topological phases of matter or quantum spin glasses to future fault-tolerant quantum computation. 

\end{abstract}

\maketitle

\section*{Introduction}
Quantum computers are expected to solve certain problems intractable in their classical counterparts. A prominent example given by Feynman is to simulate quantum many-body problems, ranging from strongly correlated materials~\cite{kennes2021moire,jafferis2022traversable} to quantum chemistry~\cite{lanyon2010towards,google2020hartree} and the dynamical evolution of quantum many-body systems~ \cite{zhang2017timecrystal,geier2021floquet}. 
Quantum simulations can be performed in a well-controlled platform with interactions native to the physical system: neutral atoms in optical lattices or optical cavities are used for models with nearest-neighbor hopping~\cite{bloch2012quantum,gross2021quantum} or all-to-all couplings~\cite{barontini2015deterministic,pedrozo2020entanglement}; whereas platforms such as trapped ions, Rydberg atoms, nitrogen-vacancy spin defects in diamonds, and molecules are well-suited for simulating long-range interacting physics~\cite{kim2010quantum,britton2012engineered,choi2017observation,kandala2017hardware}. The quest for higher flexibility and programmability has attracted much attention recently, where the interactions are encoded beyond these regimes. Qubits interacting on non-Euclidean geometries~\cite{kollar2019hyperbolic,chen2023hyperbolic} can have eigenenergies and correlations different from their Euclidean counterparts, with a prominent motivation to simulate physics in models of quantum gravity~\cite{ambjorn2004emergence}. While in solid-state systems such as superconducting qubits, the interactions are hard-wired by design~\cite{kollar2019hyperbolic,chen2023hyperbolic}, in atomic systems such as cavity quantum electrodynamics (c-QED), new modulation schemes have shown powerful tunability across different types of graphs without the need to physically re-wire~\cite{bentsen2019treelike,periwal2021programmable}.

Trapped ion systems afford such programmability: collective phonon modes mediate the interactions, and modulated laser drive fields can generate tunable non-local couplings. Quantum simulation experiments performed so far have realized long-range approximate power-law decaying interactions generated in the ``dispersive coupling`` regime, where global laser fields couple to all phonon modes\cite{zhang2017observation,joshi2022observing}. Though successful in demonstrating non-equilibrium dynamics hard to tract by classical computers~\cite{zhang2017observation}, the limited tunability (finite interaction range) and the residual spin-phonon coupling limits future scalability~\cite{monroe2021programmable}. 
In recent years, new theory proposals have emerged to engineer coupling graphs such as nearest-neighbor interactions in 1D and 2D systems~\cite{rajabi2019dynamical,teoh2020machine,shapira2020theory,manovitz2020quantum}, followed by experimental demonstrations of time-reversal-breaking dynamics in triangular ladders~\cite{shapira2023quantum}.

\begin{figure*}[!t]
    \centering
    \includegraphics[width = \textwidth]{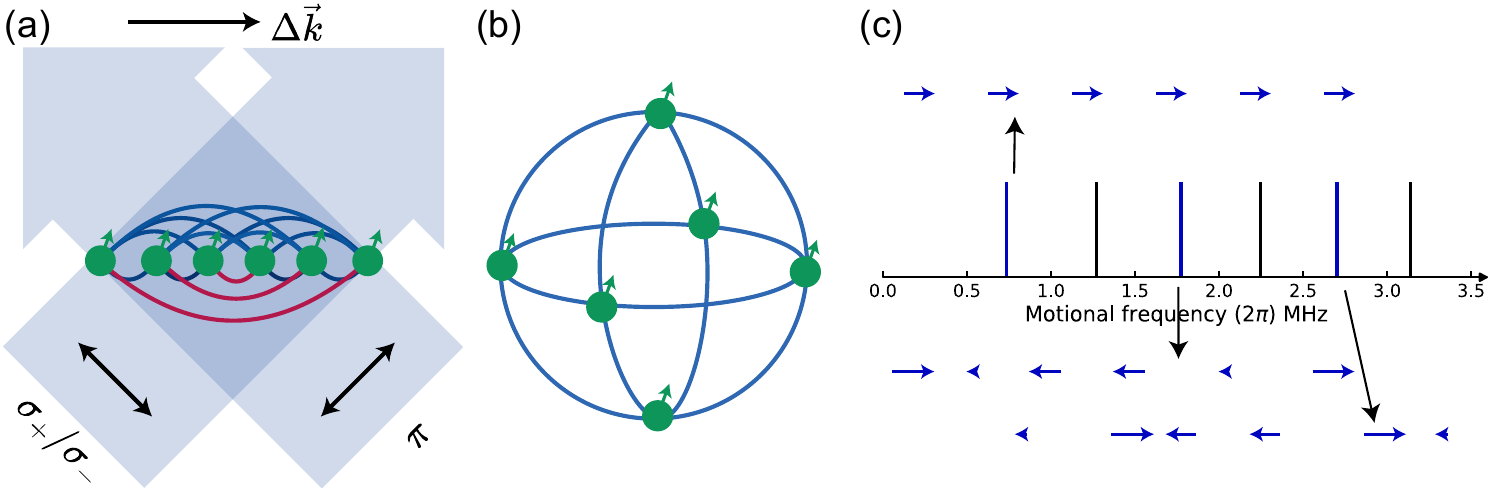}
    \caption{
     Illustration of the trapped-ion hybrid quantum information processor. (a) A chain of $\mathrm{^9 Be^+}$ ions are confined in a RF Paul trap in a linear configuration. A pair of Raman beams provides beatnote frequencies symmetrically detuned from the qubit transition $\omega_0 \pm \mu$, mediating the spin-spin interactions. The blue (red) curves denote near equal (zero) coupling strength between the connected ion pairs. (b) With 6 spins, the interaction illustrated in (a) can be mapped to nearest-neighbor interaction on a sphere. (c) These interactions are generated through a weighted combination of coupling to modes 1,3,5 (blue solid lines). The blue arrows visualize the corresponding normal mode eigenvectors. } 
    
    \label{fig:illustration}
\end{figure*}

In this work, we experimentally demonstrate, for the first time to our knowledge, that global laser fields can be modulated to encode qubits in high-dimensional non-Euclidean spaces, exemplified by up to 8 qubits interacting on a four-dimensional hyper-sphere, an example of positively-curved elliptical space.  
We study the temporal dynamics of many-body quantum systems out of equilibrium without intrinsic spin-phonon errors under programmable geometries. We experimentally tune between commuting Ising-type and non-commuting Heisenberg-type Hamiltonian models with tunable coupling strengths and durations and study the non-equilibrium dynamics with extensions to Floquet models~\cite{geier2021floquet,scholl2022microwave}. Our hybrid analog-digital quantum simulation method offers unique parallelism, thus complementing fully digital quantum computers~\cite{Debnath2016,Honeywell_QCCD}.

 \section*{Tunable coupling graphs by frequency-domain modulation}
 The non-local Coulomb coupling in principle allows arbitrary connectivity in a trapped ion system, but requires individual addressing and classical control techniques to entangle the desired spin states without excess phonon occupation~\cite{Debnath2016}. With global laser fields,  previous experiments have focused on either collectively coupling to all modes~\cite{zhang2017observation,joshi2022observing} or tuning the frequency to near the center-of-mass mode~\cite{martinez2016real,britton2012engineered}, while novel theories propose the combination of selected phonon modes for engineering tunable graphs~\cite{shapira2020theory}. Here, we modulate the global laser fields in the frequency domain, where the spin-dependent displacements of the ions can be decomposed into couplings with different normal motional modes, forming different correlation patterns. By predominantly coupling to a single mode, the phonon-mediated interaction inherits a structure from the normal mode eigenvectors (Fig.~\ref{fig:illustration}). The spin-spin interactions can be represented by a non-local weighted graph, where the weights are generated by a linear combination of the coupling strength to the individual modes, yielding tunable couplings with $(N-1)$ degrees of freedom in an $N$-ion chain~\cite{shapira2020theory}.

Here, we focus on engineering qubit interconnections beyond the physical dimension. A key to our scheme is to reverse the sign of unwanted interaction between certain ion pairs when coupling to certain modes, and by linking together the global interactions, the average interactions between these pairs vanish, forming connectivities for the desired geometries. For example, Fig.\ref{fig:illustration}(c) illustrates that six qubits on a sphere can be encoded through coupling to the first, third, and fifth axial modes with appropriate weights. Moreover, we can change the coupling patterns and explore a more general class of unique interactions by modifying the motional spectrum of the ion chain, such as generating tree-like geometries and qubits on a torus under an anharmonic trapping potential~\cite{wu2022Hamiltoniansupplemental}. 

\begin{figure*}[!t]
    \centering
    \includegraphics[width = \textwidth]{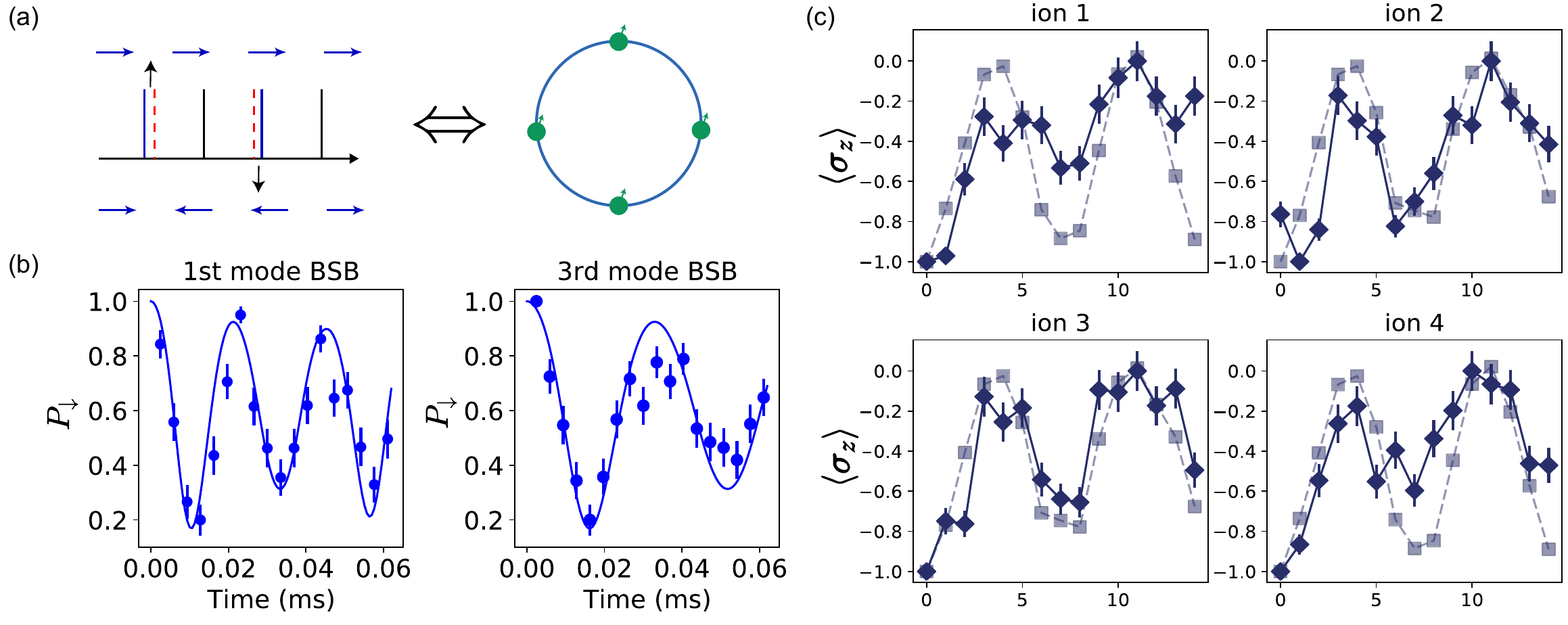}
    \caption{Illustration of qubit interaction on a plaquette with four ions. (a) The interaction is generated with equal participation of the first and third modes.  
    (b) Calibration of the BSB time evolution on the modes used in Hamiltonian engineering. Blue dots denote the measurements of the average population on the $\ket{\downarrow}$ state and blue solid lines are numerical simulations of BSB floppings, assuming $\bar{n} = 0.1$. (c) 
    Time evolution of 4 qubit interaction on a plaquette.
    The measurement of site-resolved spin magnetizations is depicted as solid lines. The numerical simulations are extrapolated through dashed lines, serving as a visual comparison to the experimental results. The error bars denote one standard deviation.   }
    
    \label{fig:highD_interaction}
\end{figure*}

\section*{Hybrid quantum information processor}

Our hybrid quantum information processor contains a linear chain of $\mathrm{^9 Be^+}$ ions confined in a radio-frequency Paul trap (Fig.~\ref{fig:illustration}(b)), held in their equilibrium positions with the Coloumb interactions balancing the external harmonic confinement. The coupled ion motions in the axial direction can be decomposed to $n$ eigenmodes, where $n$ is the ion number. We encode the qubits in the ground state hyperfine levels: $\ket{\downarrow}_z\equiv~^2 S_{1/2} \ket{F=2,m_F=2}$ and $\ket{\uparrow}_z\equiv~^2 S_{1/2} \ket{F=1,m_F=1}$, with a frequency difference of $\sim$1.018~GHz under a magnetic field of $\sim$120~G. 
 The experiment begins by cooling all the motional modes of the ions to near the ground state using continuous Raman sideband cooling~\cite{wu2023continuous}. We calibrate the spin-motion coupling by driving the blue sideband (BSB) transitions on the desired modes used for mediating the Ising interaction. Fig.~\ref{fig:highD_interaction}(b)) shows the BSB calibration in a 4-ion chain. We apply a pair of global Raman laser beams, one of which is bi-chromatic such that the beat-note frequencies are $\omega_0 \pm \mu$, symmetrically detuned from the qubit transition $\omega_0$. One of the beams is linear polarized ($\pi$), the other has linear polarization (equal $\sigma+$/$\sigma-$)  plus a small non-zero term circular polarized term to suppress the Stark shift to less than 500 Hz~\cite{wu2023quantum}, limited by the Zeeman-sensitive qubit frequency drifts under fluctuating magnetic fields. The Raman beatnote generates a spin-dependent optical dipole force via coupling to the axial motions of the ions, which mediates the spin-spin interaction~\cite{porras2004effective}. 
We choose axial modes because of their near-even spacings well-separated in the frequency domain, easing frequency modulation schemes~\cite{shapira2020theory}. We vary the detunings over the mode spectrum so that positive and negative coupling terms could exist simultaneously~\cite {wu2022Hamiltoniansupplemental}. When the laser detuning $\mu$ is close to one specific motional mode frequency $\omega_M$ with $|\mu -\omega_M|\ll  | \mu - \omega_m|_{M\neq m}$, we generate the mode-specific Ising interaction with a predominant coupling to one mode:

\begin{equation}
     H_M = \sum_{i<j} J_{ij}(\mu_M)\sigma_i^\phi\sigma_j^\phi,
     \label{Eqn:Mattis_model}
 \end{equation}
where $J_{ij}(\mu_M)$ is the spin-spin interaction term when coupling to mode $\omega_M$ and $J_{ij}(\mu_M)\propto b_{i,M}b_{j,M}/(\mu_M -\omega_M)$. $b_{i,M}$ is the eigenvector component between the  $i$-th ion and mode M, $\sigma_i^\phi = \sigma_i^x\cos{\phi} -\sigma_i^y\sin{\phi}$ is the Pauli operator for the $i$-th ion with a tunable phase $\phi$, and the reduced Planck’s constant $\hbar$ is set to 1. In previous quantum simulation experiments~\cite{monroe2021programmable}, the spin-phonon coupling term is neglected in the dispersive regime when the detuning $|\mu -\omega_m|\gg \eta_{i,m}\Omega_i$ at the cost of residual errors~\cite{kim2010quantum}. However, by making use of the axial mode structure, we mitigate this effect by choosing the detunings to our desired coupling mode $\delta_M= \mu - \omega_M$ to be a common divisor to detuning of all other modes
$\delta_m = \mu - \omega_m= k\delta_M (k\in \mathbb{Z}) $. An additional benefit is a speed-up of entangling gates compared to the adiabatic condition~\cite{schafer2018fast}. 


We first choose the appropriate laser phase to generate the $\sigma_i^x\sigma_j^x$ type Hamiltonian. We sequentially apply Ising interactions $H_a, H_b,..., H_N$ with tunable time duration of $\tau_a, \tau_b,...,\tau_N$, where every $H_m$ generates a $\sigma_x\sigma_x$ interaction with different weights by coupling to the $m$-th mode (Eq.\ref{Eqn:Mattis_model}). In the case of all commuting terms, we create the following effective Hamiltonian:
\begin{eqnarray}
{H_{eff}}&& =\frac{\sum_{m} H_m \tau_m}{\sum_m \tau_m} = \frac{\sum_{i<j} \sum_{m}  J_{ij}(\mu_m) \tau_m }{\sum_m \tau_m} \sigma_i^x\sigma_j^x\nonumber\\
&&=\sum_{i<j}\bar{J_{ij}}\sigma_i^x\sigma_j^x,
\end{eqnarray}

 where we denote $\bar{J_{ij}}$ to be the effective coupling strength and  $\bar{J_{ij}}\propto\sum_{m} b_{i,m}b_{j,m}\tau_M/(\mu_m -\omega_m) $.
Through concatenating continuously tunable (analog) blocks with different laser detunings, the system is subject to a programmable quantum evolution characterized by $\bar{J_{ij}}$. 
Such couplings correspond to the nearest-neighbor interactions in plaquette, sphere, and hypersphere for 4, 6, and 8 ions, respectively.  At the end of each global interaction layer, we perform single-site readouts of the qubit magnetizations by illuminating the ion chain with resonant detection light and collecting the spin-dependent fluorescence. 


\begin{figure}[!t]
    \centering
    \includegraphics[width = 0.5\textwidth]{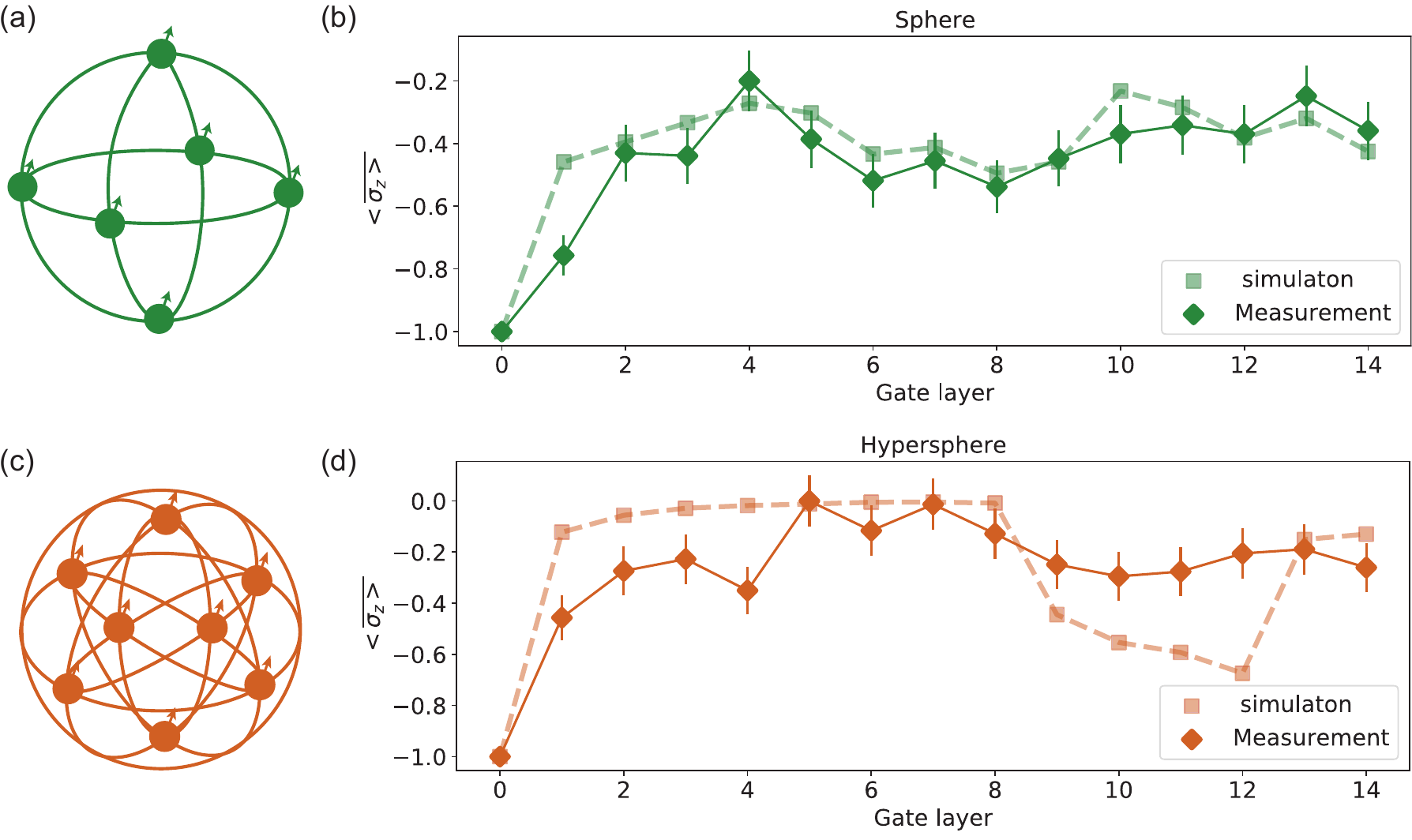}
    \caption{Realization of Ising interaction in a higher dimension sphere. 
   (a)(c) Illustration of qubit interaction of nearest-neighbor interaction on a sphere and hypersphere, with 6 and 8 qubits, respectively. 
   (b)(d) Measurement of the average spin magnetization $\overline{\langle\sigma_z\rangle}$. Solid lines extrapolate the spin magnetizations measured in the experiment, and solid rhombuses are the experimental data. Dash lines extrapolate the numerical calculations as a side-by-side comparison to the experimental result. Error bars denote one standard deviation.}
    
    \label{fig:6&8ions}
\end{figure}

\section*{Interaction in higher-dimension with 1D chain}

We first show that a 4-ion chain can be encoded in the smallest setting of 2D nearest-neighbor interactions in a square plaquette (Fig.~\ref{fig:highD_interaction}(a)). We prepare the ions on the $\ket{\downarrow}_{z, N}$ state and apply successive $H_1$ and $H_3$ interactions iteratively. $H_1$ couples to the first mode generating all-to-all couplings $J_{ij}= \mathcal{J}$ for all $i,j$ and $H_3$ couples to the third mode with reversed coupling in two of the pairs $J_{14},J_{23} = -\mathcal{J}$. The laser detunings of these two interactions are $\delta_1/2\pi = 107.6$~kHz from the first mode and $\delta_3/2\pi = -71.14$~kHz from the third mode, with interaction time $\tau_{1} = 2\pi/\delta_{1} = 9.29~\mu s$ and $\tau_{3} = 2\pi/\delta_{3} = 14.06~\mu s$. After the concatenated unitary evolution, the effective Hamiltonian generates a 2D nearest-neighbor
interaction, which we measure via site-resolve spin magnetization $\langle\sigma_z\rangle$ (Fig.~\ref{fig:highD_interaction}(c)). The understand the deviations of the experimental results from the implemented Hamiltonian,  we perform numerical simulations of the implemented Hamiltonian taking into account systematic experimental offsets, including qubit frequency and laser intensity gradient\cite{wu2022Hamiltoniansupplemental}, and find reasonable agreements between the simulation and the experimental data. 

We next generalize the method to higher dimensional spaces. With six ions, a minimum graph corresponding to a sphere with evenly distributed qubits (two on the north and south poles, and four evenly distributed on the equator) can be generated by removing selected vertices (red lines in Fig.~\ref{fig:illustration}(a)), which is similar to removing the ``cross" vertices in our previous case of 4 qubits. As a primary example of non-Euclidean space with positive curvature, we can extend this to 8 qubits and encode them on a 4-sphere. The protocols are periodic drives applied to the system with coupling to the odd number modes, namely modes 1,3,5 in the 6-ion chain and modes 1,3,5,7 in the 8-ion chain. 
The detunings from the modes and layer durations can be found in~\cite{wu2022Hamiltoniansupplemental}. Similarly, we capture the temporal dynamics of the system by measuring spin magnetizations. Fig.~\ref{fig:6&8ions}(b)(c) show the average spin dynamics as a function of the gate layers applied. Our results are consistent with numerical calculation, considering the actual Hamiltonian with the experimental offset (see S.I.~\cite{wu2022Hamiltoniansupplemental}).

\section*{Floquet XY and Heisenberg models}

\begin{figure}[!t]
    \centering
    \includegraphics[width = 0.5\textwidth]{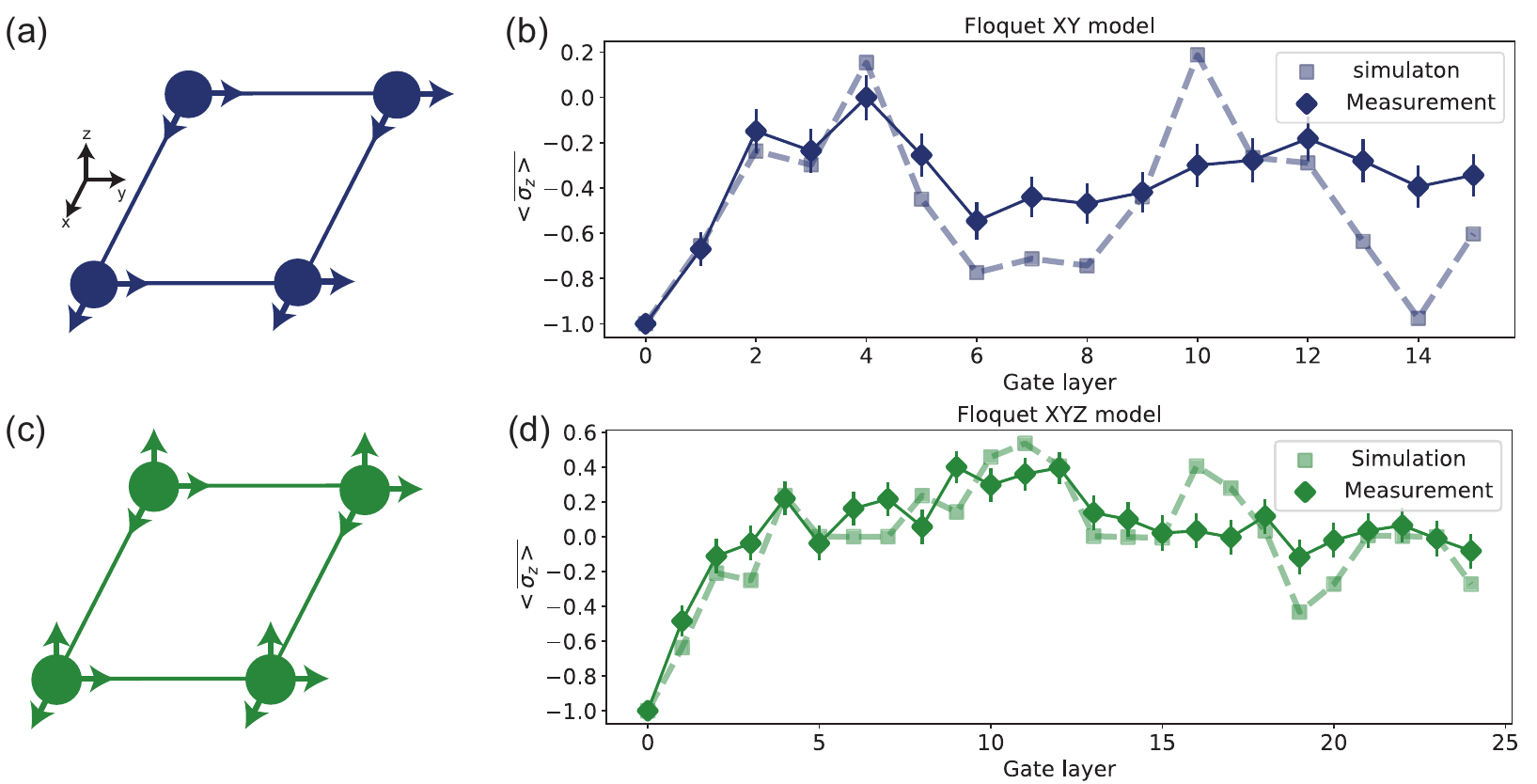}
    \caption{Realization of Floquet XY and XYZ Hamiltonians. Illustration of (a)XY (c) Heisenberg XYZ model in a 2D nearest neighbor interaction with four ions. The (b) navy and (d) green rhombuses denote the evolution of the $\langle\sigma_z\rangle$ as a function of the applied analog blocks under an isotropic (b) XY (d) XYZ 2D nearest neighbor interaction with four ions. The error bars denote one standard deviation. The dashed lines extrapolate the numerical simulation of the ideal quantum evolution, considering the experimental offsets.  }
    
    \label{fig:4ions_Floquet}
\end{figure}

 The flexible control of layers of global interaction makes our schemes favorable to extend to other spin degrees of freedom and even non-commuting spin-spin interactions, which native Hamiltonians do not exist in our platform. As an example, we  implement the same 2D interaction on a plaquette on a Floquet XY model by applying a periodic drive with $T = 2(t_1+t_3)$ to the system, which consists of four successive global interaction layers:
\begin{equation}
    H=\left\{\begin{array}{l}H_1^{XX}=\sum_{i<j} J_{i j}^{(1)} \sigma_i^x \sigma_j^x \qquad t= t_1\\ H_3^{XX}=\sum_{i<j} J_{i j}^{(3)} \sigma_i^x \sigma_j^x \qquad t= t_3 \\ H_1^{YY}=\sum_{i<j} J_{i j}^{(1)} \sigma_i^y \sigma_j^y  \qquad t= t_1\\
    H_3^{YY}=\sum_{i<j} J_{i j}^{(3)} \sigma_i^y \sigma_j^y  \qquad t= t_3\end{array}\right..
\end{equation}

Here, $H_1$ and $H_3$ are the two types of interaction patterns when coupled to the first and third modes. A static quantum XY model $H_{XY} = J_{ij}(\sigma_i^x \sigma_j^x+\sigma_i^y \sigma_j^y)$ has no dynamics for the ferromagnetic chain when the spins are initialized in $\ket{\downarrow\downarrow\downarrow\downarrow}_z$ state (paramagnetic spin chain). However, the system exhibits strong time-dependent dynamics under a Floquet periodic drive with suitable interaction strength of each layer. Fig.~\ref{fig:4ions_Floquet}(a) shows the experimental result of the emerging dynamics of the collective spin magnetization. We find decent agreement with the numerical model considering experimental offsets.

Furthermore, we can engineer Heisenberg Hamiltonians with tunable anisotropy with periodic modulation. We present an experimental implementation of the Heisenberg model to the same qubit interaction graph with four ions. To generate the $\sigma_z\sigma_z$-type interaction, we drive the system into a spin-rotating frame using a global $\pi/2$ y-rotation, followed by $\sigma_x\sigma_x$ global interaction, and then a global $-\pi/2$ y-rotation. By that, a trotter step consists of 8 global operations\cite{arrazola2016digital}:

\begin{eqnarray}
U_{XYZ}(t)=&e^{-i H_{1}^{XX }t_1}e^{-i H_{3}^{XX }t_3}e^{-i H_{1}^{YY }t_1}e^{-i H_{3}^{YY }t_3}\nonumber\\
&R_y(
\pi/4) e^{-i H_{1}^{XX }t_1}e^{-i H_{3}^{XX }t_3} R_y^{\dagger}(
\pi/4).
\end{eqnarray}

Fig.~\ref{fig:4ions_Floquet} (b) shows the experimental result of the emerging dynamics of the collective spin magnetization under the Floquet Heisenberg model. We compare it with numerical simulations, validating that our protocol implements the target Hamiltonian after up to 24 analog blocks. 

\section*{Conclusion and outlook}
We have experimentally realized qubit interactions in a higher dimensional space on a 1D ion chain of up to 8 qubits using the analog-digital hybrid approach with global laser beams. Furthermore, we apply this protocol to generate Floquet XY and Heisenberg models of the 2D nearest neighbor interaction using up to 24 global interaction blocks with continuously tunable parameters. 
Our results contribute to the engineering of multi-dimension Hamiltonians using global interactions with analog-digital hybrid protocols, which opens up the full potential of the trapped ion quantum simulator beyond the long-range interaction regime, laying the foundations for more complex geometries~\cite{manovitz2020quantum}. 
Combining our method with single-spin addressing and optical tweezers~\cite{olsacher2020scalable}, generating arbitrary Ising graphs at a lower engineering cost than a universal quantum computer will be possible. This new platform will allow us to perform quantum simulation of quantum systems with complex geometries such as MaxCut problems\cite{ebadi2022quantum}, quantum spin glasses\cite{gopalakrishnan2011frustration} and the 2D Kagome lattice\cite{korenblit2012quantum}. Furthermore, our scheme might inspire the direct encoding of a quantum error correction circuit~\cite{Honeywell_QEC}, with potential advantages in using non-Euclidean geometries~\cite{kitaev2003fault}.
\\

\begin{acknowledgements}
We thank Luis Orozco, Dries Sels, Wenchao Ge, Ye Wang, and Zijian Ding for helpful discussions and critical reading of the manuscript. This work was performed at New York University, with financial support from the US Department of Energy.

\end{acknowledgements}

\bibliography{Hamiltonian_engineering_reference}

\begin{thebibliography}{43}
\expandafter\ifx\csname natexlab\endcsname\relax\def\natexlab#1{#1}\fi
\expandafter\ifx\csname bibnamefont\endcsname\relax
  \def\bibnamefont#1{#1}\fi
\expandafter\ifx\csname bibfnamefont\endcsname\relax
  \def\bibfnamefont#1{#1}\fi
\expandafter\ifx\csname citenamefont\endcsname\relax
  \def\citenamefont#1{#1}\fi
\expandafter\ifx\csname url\endcsname\relax
  \def\url#1{\texttt{#1}}\fi
\expandafter\ifx\csname urlprefix\endcsname\relax\def\urlprefix{URL }\fi
\providecommand{\bibinfo}[2]{#2}
\providecommand{\eprint}[2][]{\url{#2}}

\bibitem[{\citenamefont{Kennes et~al.}(2021)\citenamefont{Kennes, Claassen,
  Xian, Georges, Millis, Hone, Dean, Basov, Pasupathy, and
  Rubio}}]{kennes2021moire}
\bibinfo{author}{\bibfnamefont{D.~M.} \bibnamefont{Kennes}},
  \bibinfo{author}{\bibfnamefont{M.}~\bibnamefont{Claassen}},
  \bibinfo{author}{\bibfnamefont{L.}~\bibnamefont{Xian}},
  \bibinfo{author}{\bibfnamefont{A.}~\bibnamefont{Georges}},
  \bibinfo{author}{\bibfnamefont{A.~J.} \bibnamefont{Millis}},
  \bibinfo{author}{\bibfnamefont{J.}~\bibnamefont{Hone}},
  \bibinfo{author}{\bibfnamefont{C.~R.} \bibnamefont{Dean}},
  \bibinfo{author}{\bibfnamefont{D.}~\bibnamefont{Basov}},
  \bibinfo{author}{\bibfnamefont{A.~N.} \bibnamefont{Pasupathy}},
  \bibnamefont{and} \bibinfo{author}{\bibfnamefont{A.}~\bibnamefont{Rubio}},
  \bibinfo{journal}{Nature Physics} \textbf{\bibinfo{volume}{17}},
  \bibinfo{pages}{155} (\bibinfo{year}{2021}).

\bibitem[{\citenamefont{Jafferis et~al.}(2022)\citenamefont{Jafferis, Zlokapa,
  Lykken, Kolchmeyer, Davis, Lauk, Neven, and
  Spiropulu}}]{jafferis2022traversable}
\bibinfo{author}{\bibfnamefont{D.}~\bibnamefont{Jafferis}},
  \bibinfo{author}{\bibfnamefont{A.}~\bibnamefont{Zlokapa}},
  \bibinfo{author}{\bibfnamefont{J.~D.} \bibnamefont{Lykken}},
  \bibinfo{author}{\bibfnamefont{D.~K.} \bibnamefont{Kolchmeyer}},
  \bibinfo{author}{\bibfnamefont{S.~I.} \bibnamefont{Davis}},
  \bibinfo{author}{\bibfnamefont{N.}~\bibnamefont{Lauk}},
  \bibinfo{author}{\bibfnamefont{H.}~\bibnamefont{Neven}}, \bibnamefont{and}
  \bibinfo{author}{\bibfnamefont{M.}~\bibnamefont{Spiropulu}},
  \bibinfo{journal}{Nature} \textbf{\bibinfo{volume}{612}}, \bibinfo{pages}{51}
  (\bibinfo{year}{2022}).

\bibitem[{\citenamefont{Lanyon et~al.}(2010)\citenamefont{Lanyon, Whitfield,
  Gillett, Goggin, Almeida, Kassal, Biamonte, Mohseni, Powell, Barbieri
  et~al.}}]{lanyon2010towards}
\bibinfo{author}{\bibfnamefont{B.~P.} \bibnamefont{Lanyon}},
  \bibinfo{author}{\bibfnamefont{J.~D.} \bibnamefont{Whitfield}},
  \bibinfo{author}{\bibfnamefont{G.~G.} \bibnamefont{Gillett}},
  \bibinfo{author}{\bibfnamefont{M.~E.} \bibnamefont{Goggin}},
  \bibinfo{author}{\bibfnamefont{M.~P.} \bibnamefont{Almeida}},
  \bibinfo{author}{\bibfnamefont{I.}~\bibnamefont{Kassal}},
  \bibinfo{author}{\bibfnamefont{J.~D.} \bibnamefont{Biamonte}},
  \bibinfo{author}{\bibfnamefont{M.}~\bibnamefont{Mohseni}},
  \bibinfo{author}{\bibfnamefont{B.~J.} \bibnamefont{Powell}},
  \bibinfo{author}{\bibfnamefont{M.}~\bibnamefont{Barbieri}},
  \bibnamefont{et~al.}, \bibinfo{journal}{Nature chemistry}
  \textbf{\bibinfo{volume}{2}}, \bibinfo{pages}{106} (\bibinfo{year}{2010}).

\bibitem[{\citenamefont{Quantum et~al.}(2020)\citenamefont{Quantum,
  Collaborators*†, Arute, Arya, Babbush, Bacon, Bardin, Barends, Boixo,
  Broughton et~al.}}]{google2020hartree}
\bibinfo{author}{\bibfnamefont{G.~A.} \bibnamefont{Quantum}},
  \bibinfo{author}{\bibnamefont{Collaborators*†}},
  \bibinfo{author}{\bibfnamefont{F.}~\bibnamefont{Arute}},
  \bibinfo{author}{\bibfnamefont{K.}~\bibnamefont{Arya}},
  \bibinfo{author}{\bibfnamefont{R.}~\bibnamefont{Babbush}},
  \bibinfo{author}{\bibfnamefont{D.}~\bibnamefont{Bacon}},
  \bibinfo{author}{\bibfnamefont{J.~C.} \bibnamefont{Bardin}},
  \bibinfo{author}{\bibfnamefont{R.}~\bibnamefont{Barends}},
  \bibinfo{author}{\bibfnamefont{S.}~\bibnamefont{Boixo}},
  \bibinfo{author}{\bibfnamefont{M.}~\bibnamefont{Broughton}},
  \bibnamefont{et~al.}, \bibinfo{journal}{Science}
  \textbf{\bibinfo{volume}{369}}, \bibinfo{pages}{1084} (\bibinfo{year}{2020}).

\bibitem[{\citenamefont{Zhang et~al.}(2017{\natexlab{a}})\citenamefont{Zhang,
  Hess, Kyprianidis, Becker, Lee, Smith, Pagano, Potirniche, Potter, Vishwanath
  et~al.}}]{zhang2017timecrystal}
\bibinfo{author}{\bibfnamefont{J.}~\bibnamefont{Zhang}},
  \bibinfo{author}{\bibfnamefont{P.~W.} \bibnamefont{Hess}},
  \bibinfo{author}{\bibfnamefont{A.}~\bibnamefont{Kyprianidis}},
  \bibinfo{author}{\bibfnamefont{P.}~\bibnamefont{Becker}},
  \bibinfo{author}{\bibfnamefont{A.}~\bibnamefont{Lee}},
  \bibinfo{author}{\bibfnamefont{J.}~\bibnamefont{Smith}},
  \bibinfo{author}{\bibfnamefont{G.}~\bibnamefont{Pagano}},
  \bibinfo{author}{\bibfnamefont{I.-D.} \bibnamefont{Potirniche}},
  \bibinfo{author}{\bibfnamefont{A.~C.} \bibnamefont{Potter}},
  \bibinfo{author}{\bibfnamefont{A.}~\bibnamefont{Vishwanath}},
  \bibnamefont{et~al.}, \bibinfo{journal}{Nature}
  \textbf{\bibinfo{volume}{543}}, \bibinfo{pages}{217}
  (\bibinfo{year}{2017}{\natexlab{a}}).

\bibitem[{\citenamefont{Geier et~al.}(2021)\citenamefont{Geier, Thaicharoen,
  Hainaut, Franz, Salzinger, Tebben, Grimshandl, Z{\"u}rn, and
  Weidem{\"u}ller}}]{geier2021floquet}
\bibinfo{author}{\bibfnamefont{S.}~\bibnamefont{Geier}},
  \bibinfo{author}{\bibfnamefont{N.}~\bibnamefont{Thaicharoen}},
  \bibinfo{author}{\bibfnamefont{C.}~\bibnamefont{Hainaut}},
  \bibinfo{author}{\bibfnamefont{T.}~\bibnamefont{Franz}},
  \bibinfo{author}{\bibfnamefont{A.}~\bibnamefont{Salzinger}},
  \bibinfo{author}{\bibfnamefont{A.}~\bibnamefont{Tebben}},
  \bibinfo{author}{\bibfnamefont{D.}~\bibnamefont{Grimshandl}},
  \bibinfo{author}{\bibfnamefont{G.}~\bibnamefont{Z{\"u}rn}}, \bibnamefont{and}
  \bibinfo{author}{\bibfnamefont{M.}~\bibnamefont{Weidem{\"u}ller}},
  \bibinfo{journal}{Science} \textbf{\bibinfo{volume}{374}},
  \bibinfo{pages}{1149} (\bibinfo{year}{2021}).

\bibitem[{\citenamefont{Bloch et~al.}(2012)\citenamefont{Bloch, Dalibard, and
  Nascimbene}}]{bloch2012quantum}
\bibinfo{author}{\bibfnamefont{I.}~\bibnamefont{Bloch}},
  \bibinfo{author}{\bibfnamefont{J.}~\bibnamefont{Dalibard}}, \bibnamefont{and}
  \bibinfo{author}{\bibfnamefont{S.}~\bibnamefont{Nascimbene}},
  \bibinfo{journal}{Nature Physics} \textbf{\bibinfo{volume}{8}},
  \bibinfo{pages}{267} (\bibinfo{year}{2012}).

\bibitem[{\citenamefont{Gross and Bakr}(2021)}]{gross2021quantum}
\bibinfo{author}{\bibfnamefont{C.}~\bibnamefont{Gross}} \bibnamefont{and}
  \bibinfo{author}{\bibfnamefont{W.~S.} \bibnamefont{Bakr}},
  \bibinfo{journal}{Nature Physics} \textbf{\bibinfo{volume}{17}},
  \bibinfo{pages}{1316} (\bibinfo{year}{2021}).

\bibitem[{\citenamefont{Barontini et~al.}(2015)\citenamefont{Barontini,
  Hohmann, Haas, Est{\`e}ve, and Reichel}}]{barontini2015deterministic}
\bibinfo{author}{\bibfnamefont{G.}~\bibnamefont{Barontini}},
  \bibinfo{author}{\bibfnamefont{L.}~\bibnamefont{Hohmann}},
  \bibinfo{author}{\bibfnamefont{F.}~\bibnamefont{Haas}},
  \bibinfo{author}{\bibfnamefont{J.}~\bibnamefont{Est{\`e}ve}},
  \bibnamefont{and} \bibinfo{author}{\bibfnamefont{J.}~\bibnamefont{Reichel}},
  \bibinfo{journal}{Science} \textbf{\bibinfo{volume}{349}},
  \bibinfo{pages}{1317} (\bibinfo{year}{2015}).

\bibitem[{\citenamefont{Pedrozo-Pe{\~n}afiel
  et~al.}(2020)\citenamefont{Pedrozo-Pe{\~n}afiel, Colombo, Shu, Adiyatullin,
  Li, Mendez, Braverman, Kawasaki, Akamatsu, Xiao
  et~al.}}]{pedrozo2020entanglement}
\bibinfo{author}{\bibfnamefont{E.}~\bibnamefont{Pedrozo-Pe{\~n}afiel}},
  \bibinfo{author}{\bibfnamefont{S.}~\bibnamefont{Colombo}},
  \bibinfo{author}{\bibfnamefont{C.}~\bibnamefont{Shu}},
  \bibinfo{author}{\bibfnamefont{A.~F.} \bibnamefont{Adiyatullin}},
  \bibinfo{author}{\bibfnamefont{Z.}~\bibnamefont{Li}},
  \bibinfo{author}{\bibfnamefont{E.}~\bibnamefont{Mendez}},
  \bibinfo{author}{\bibfnamefont{B.}~\bibnamefont{Braverman}},
  \bibinfo{author}{\bibfnamefont{A.}~\bibnamefont{Kawasaki}},
  \bibinfo{author}{\bibfnamefont{D.}~\bibnamefont{Akamatsu}},
  \bibinfo{author}{\bibfnamefont{Y.}~\bibnamefont{Xiao}}, \bibnamefont{et~al.},
  \bibinfo{journal}{Nature} \textbf{\bibinfo{volume}{588}},
  \bibinfo{pages}{414} (\bibinfo{year}{2020}).

\bibitem[{\citenamefont{Kim et~al.}(2010)\citenamefont{Kim, Chang, Korenblit,
  Islam, Edwards, Freericks, Lin, Duan, and Monroe}}]{kim2010quantum}
\bibinfo{author}{\bibfnamefont{K.}~\bibnamefont{Kim}},
  \bibinfo{author}{\bibfnamefont{M.-S.} \bibnamefont{Chang}},
  \bibinfo{author}{\bibfnamefont{S.}~\bibnamefont{Korenblit}},
  \bibinfo{author}{\bibfnamefont{R.}~\bibnamefont{Islam}},
  \bibinfo{author}{\bibfnamefont{E.~E.} \bibnamefont{Edwards}},
  \bibinfo{author}{\bibfnamefont{J.~K.} \bibnamefont{Freericks}},
  \bibinfo{author}{\bibfnamefont{G.-D.} \bibnamefont{Lin}},
  \bibinfo{author}{\bibfnamefont{L.-M.} \bibnamefont{Duan}}, \bibnamefont{and}
  \bibinfo{author}{\bibfnamefont{C.}~\bibnamefont{Monroe}},
  \bibinfo{journal}{Nature} \textbf{\bibinfo{volume}{465}},
  \bibinfo{pages}{590} (\bibinfo{year}{2010}).

\bibitem[{\citenamefont{Britton et~al.}(2012)\citenamefont{Britton, Sawyer,
  Keith, Wang, Freericks, Uys, Biercuk, and Bollinger}}]{britton2012engineered}
\bibinfo{author}{\bibfnamefont{J.~W.} \bibnamefont{Britton}},
  \bibinfo{author}{\bibfnamefont{B.~C.} \bibnamefont{Sawyer}},
  \bibinfo{author}{\bibfnamefont{A.~C.} \bibnamefont{Keith}},
  \bibinfo{author}{\bibfnamefont{C.-C.~J.} \bibnamefont{Wang}},
  \bibinfo{author}{\bibfnamefont{J.~K.} \bibnamefont{Freericks}},
  \bibinfo{author}{\bibfnamefont{H.}~\bibnamefont{Uys}},
  \bibinfo{author}{\bibfnamefont{M.~J.} \bibnamefont{Biercuk}},
  \bibnamefont{and} \bibinfo{author}{\bibfnamefont{J.~J.}
  \bibnamefont{Bollinger}}, \bibinfo{journal}{Nature}
  \textbf{\bibinfo{volume}{484}}, \bibinfo{pages}{489} (\bibinfo{year}{2012}).

\bibitem[{\citenamefont{Choi et~al.}(2017)\citenamefont{Choi, Choi, Landig,
  Kucsko, Zhou, Isoya, Jelezko, Onoda, Sumiya, Khemani
  et~al.}}]{choi2017observation}
\bibinfo{author}{\bibfnamefont{S.}~\bibnamefont{Choi}},
  \bibinfo{author}{\bibfnamefont{J.}~\bibnamefont{Choi}},
  \bibinfo{author}{\bibfnamefont{R.}~\bibnamefont{Landig}},
  \bibinfo{author}{\bibfnamefont{G.}~\bibnamefont{Kucsko}},
  \bibinfo{author}{\bibfnamefont{H.}~\bibnamefont{Zhou}},
  \bibinfo{author}{\bibfnamefont{J.}~\bibnamefont{Isoya}},
  \bibinfo{author}{\bibfnamefont{F.}~\bibnamefont{Jelezko}},
  \bibinfo{author}{\bibfnamefont{S.}~\bibnamefont{Onoda}},
  \bibinfo{author}{\bibfnamefont{H.}~\bibnamefont{Sumiya}},
  \bibinfo{author}{\bibfnamefont{V.}~\bibnamefont{Khemani}},
  \bibnamefont{et~al.}, \bibinfo{journal}{Nature}
  \textbf{\bibinfo{volume}{543}}, \bibinfo{pages}{221} (\bibinfo{year}{2017}).

\bibitem[{\citenamefont{Kandala et~al.}(2017)\citenamefont{Kandala, Mezzacapo,
  Temme, Takita, Brink, Chow, and Gambetta}}]{kandala2017hardware}
\bibinfo{author}{\bibfnamefont{A.}~\bibnamefont{Kandala}},
  \bibinfo{author}{\bibfnamefont{A.}~\bibnamefont{Mezzacapo}},
  \bibinfo{author}{\bibfnamefont{K.}~\bibnamefont{Temme}},
  \bibinfo{author}{\bibfnamefont{M.}~\bibnamefont{Takita}},
  \bibinfo{author}{\bibfnamefont{M.}~\bibnamefont{Brink}},
  \bibinfo{author}{\bibfnamefont{J.~M.} \bibnamefont{Chow}}, \bibnamefont{and}
  \bibinfo{author}{\bibfnamefont{J.~M.} \bibnamefont{Gambetta}},
  \bibinfo{journal}{Nature} \textbf{\bibinfo{volume}{549}},
  \bibinfo{pages}{242} (\bibinfo{year}{2017}).

\bibitem[{\citenamefont{Koll{\'a}r et~al.}(2019)\citenamefont{Koll{\'a}r,
  Fitzpatrick, and Houck}}]{kollar2019hyperbolic}
\bibinfo{author}{\bibfnamefont{A.~J.} \bibnamefont{Koll{\'a}r}},
  \bibinfo{author}{\bibfnamefont{M.}~\bibnamefont{Fitzpatrick}},
  \bibnamefont{and} \bibinfo{author}{\bibfnamefont{A.~A.} \bibnamefont{Houck}},
  \bibinfo{journal}{Nature} \textbf{\bibinfo{volume}{571}}, \bibinfo{pages}{45}
  (\bibinfo{year}{2019}).

\bibitem[{\citenamefont{Chen et~al.}(2023)\citenamefont{Chen, Brand, Helbig,
  Hofmann, Imhof, Fritzsche, Kie{\ss}ling, Stegmaier, Upreti, Neupert
  et~al.}}]{chen2023hyperbolic}
\bibinfo{author}{\bibfnamefont{A.}~\bibnamefont{Chen}},
  \bibinfo{author}{\bibfnamefont{H.}~\bibnamefont{Brand}},
  \bibinfo{author}{\bibfnamefont{T.}~\bibnamefont{Helbig}},
  \bibinfo{author}{\bibfnamefont{T.}~\bibnamefont{Hofmann}},
  \bibinfo{author}{\bibfnamefont{S.}~\bibnamefont{Imhof}},
  \bibinfo{author}{\bibfnamefont{A.}~\bibnamefont{Fritzsche}},
  \bibinfo{author}{\bibfnamefont{T.}~\bibnamefont{Kie{\ss}ling}},
  \bibinfo{author}{\bibfnamefont{A.}~\bibnamefont{Stegmaier}},
  \bibinfo{author}{\bibfnamefont{L.~K.} \bibnamefont{Upreti}},
  \bibinfo{author}{\bibfnamefont{T.}~\bibnamefont{Neupert}},
  \bibnamefont{et~al.}, \bibinfo{journal}{Nature Communications}
  \textbf{\bibinfo{volume}{14}}, \bibinfo{pages}{622} (\bibinfo{year}{2023}).

\bibitem[{\citenamefont{Ambj{\o}rn et~al.}(2004)\citenamefont{Ambj{\o}rn,
  Jurkiewicz, and Loll}}]{ambjorn2004emergence}
\bibinfo{author}{\bibfnamefont{J.}~\bibnamefont{Ambj{\o}rn}},
  \bibinfo{author}{\bibfnamefont{J.}~\bibnamefont{Jurkiewicz}},
  \bibnamefont{and} \bibinfo{author}{\bibfnamefont{R.}~\bibnamefont{Loll}},
  \bibinfo{journal}{Physical review letters} \textbf{\bibinfo{volume}{93}},
  \bibinfo{pages}{131301} (\bibinfo{year}{2004}).

\bibitem[{\citenamefont{Bentsen et~al.}(2019)\citenamefont{Bentsen, Hashizume,
  Buyskikh, Davis, Daley, Gubser, and Schleier-Smith}}]{bentsen2019treelike}
\bibinfo{author}{\bibfnamefont{G.}~\bibnamefont{Bentsen}},
  \bibinfo{author}{\bibfnamefont{T.}~\bibnamefont{Hashizume}},
  \bibinfo{author}{\bibfnamefont{A.~S.} \bibnamefont{Buyskikh}},
  \bibinfo{author}{\bibfnamefont{E.~J.} \bibnamefont{Davis}},
  \bibinfo{author}{\bibfnamefont{A.~J.} \bibnamefont{Daley}},
  \bibinfo{author}{\bibfnamefont{S.~S.} \bibnamefont{Gubser}},
  \bibnamefont{and}
  \bibinfo{author}{\bibfnamefont{M.}~\bibnamefont{Schleier-Smith}},
  \bibinfo{journal}{Physical review letters} \textbf{\bibinfo{volume}{123}},
  \bibinfo{pages}{130601} (\bibinfo{year}{2019}).

\bibitem[{\citenamefont{Periwal et~al.}(2021)\citenamefont{Periwal, Cooper,
  Kunkel, Wienand, Davis, and Schleier-Smith}}]{periwal2021programmable}
\bibinfo{author}{\bibfnamefont{A.}~\bibnamefont{Periwal}},
  \bibinfo{author}{\bibfnamefont{E.~S.} \bibnamefont{Cooper}},
  \bibinfo{author}{\bibfnamefont{P.}~\bibnamefont{Kunkel}},
  \bibinfo{author}{\bibfnamefont{J.~F.} \bibnamefont{Wienand}},
  \bibinfo{author}{\bibfnamefont{E.~J.} \bibnamefont{Davis}}, \bibnamefont{and}
  \bibinfo{author}{\bibfnamefont{M.}~\bibnamefont{Schleier-Smith}},
  \bibinfo{journal}{Nature} \textbf{\bibinfo{volume}{600}},
  \bibinfo{pages}{630} (\bibinfo{year}{2021}).

\bibitem[{\citenamefont{Zhang et~al.}(2017{\natexlab{b}})\citenamefont{Zhang,
  Pagano, Hess, Kyprianidis, Becker, Kaplan, Gorshkov, Gong, and
  Monroe}}]{zhang2017observation}
\bibinfo{author}{\bibfnamefont{J.}~\bibnamefont{Zhang}},
  \bibinfo{author}{\bibfnamefont{G.}~\bibnamefont{Pagano}},
  \bibinfo{author}{\bibfnamefont{P.~W.} \bibnamefont{Hess}},
  \bibinfo{author}{\bibfnamefont{A.}~\bibnamefont{Kyprianidis}},
  \bibinfo{author}{\bibfnamefont{P.}~\bibnamefont{Becker}},
  \bibinfo{author}{\bibfnamefont{H.}~\bibnamefont{Kaplan}},
  \bibinfo{author}{\bibfnamefont{A.~V.} \bibnamefont{Gorshkov}},
  \bibinfo{author}{\bibfnamefont{Z.-X.} \bibnamefont{Gong}}, \bibnamefont{and}
  \bibinfo{author}{\bibfnamefont{C.}~\bibnamefont{Monroe}},
  \bibinfo{journal}{Nature} \textbf{\bibinfo{volume}{551}},
  \bibinfo{pages}{601} (\bibinfo{year}{2017}{\natexlab{b}}).

\bibitem[{\citenamefont{Joshi et~al.}(2022)\citenamefont{Joshi, Kranzl,
  Schuckert, Lovas, Maier, Blatt, Knap, and Roos}}]{joshi2022observing}
\bibinfo{author}{\bibfnamefont{M.~K.} \bibnamefont{Joshi}},
  \bibinfo{author}{\bibfnamefont{F.}~\bibnamefont{Kranzl}},
  \bibinfo{author}{\bibfnamefont{A.}~\bibnamefont{Schuckert}},
  \bibinfo{author}{\bibfnamefont{I.}~\bibnamefont{Lovas}},
  \bibinfo{author}{\bibfnamefont{C.}~\bibnamefont{Maier}},
  \bibinfo{author}{\bibfnamefont{R.}~\bibnamefont{Blatt}},
  \bibinfo{author}{\bibfnamefont{M.}~\bibnamefont{Knap}}, \bibnamefont{and}
  \bibinfo{author}{\bibfnamefont{C.~F.} \bibnamefont{Roos}},
  \bibinfo{journal}{Science} \textbf{\bibinfo{volume}{376}},
  \bibinfo{pages}{720} (\bibinfo{year}{2022}).

\bibitem[{\citenamefont{Monroe et~al.}(2021)\citenamefont{Monroe, Campbell,
  Duan, Gong, Gorshkov, Hess, Islam, Kim, Linke, Pagano
  et~al.}}]{monroe2021programmable}
\bibinfo{author}{\bibfnamefont{C.}~\bibnamefont{Monroe}},
  \bibinfo{author}{\bibfnamefont{W.~C.} \bibnamefont{Campbell}},
  \bibinfo{author}{\bibfnamefont{L.-M.} \bibnamefont{Duan}},
  \bibinfo{author}{\bibfnamefont{Z.-X.} \bibnamefont{Gong}},
  \bibinfo{author}{\bibfnamefont{A.~V.} \bibnamefont{Gorshkov}},
  \bibinfo{author}{\bibfnamefont{P.}~\bibnamefont{Hess}},
  \bibinfo{author}{\bibfnamefont{R.}~\bibnamefont{Islam}},
  \bibinfo{author}{\bibfnamefont{K.}~\bibnamefont{Kim}},
  \bibinfo{author}{\bibfnamefont{N.~M.} \bibnamefont{Linke}},
  \bibinfo{author}{\bibfnamefont{G.}~\bibnamefont{Pagano}},
  \bibnamefont{et~al.}, \bibinfo{journal}{Reviews of Modern Physics}
  \textbf{\bibinfo{volume}{93}}, \bibinfo{pages}{025001}
  (\bibinfo{year}{2021}).

\bibitem[{\citenamefont{Rajabi et~al.}(2019)\citenamefont{Rajabi, Motlakunta,
  Shih, Kotibhaskar, Quraishi, Ajoy, and Islam}}]{rajabi2019dynamical}
\bibinfo{author}{\bibfnamefont{F.}~\bibnamefont{Rajabi}},
  \bibinfo{author}{\bibfnamefont{S.}~\bibnamefont{Motlakunta}},
  \bibinfo{author}{\bibfnamefont{C.-Y.} \bibnamefont{Shih}},
  \bibinfo{author}{\bibfnamefont{N.}~\bibnamefont{Kotibhaskar}},
  \bibinfo{author}{\bibfnamefont{Q.}~\bibnamefont{Quraishi}},
  \bibinfo{author}{\bibfnamefont{A.}~\bibnamefont{Ajoy}}, \bibnamefont{and}
  \bibinfo{author}{\bibfnamefont{R.}~\bibnamefont{Islam}},
  \bibinfo{journal}{npj Quantum Information} \textbf{\bibinfo{volume}{5}},
  \bibinfo{pages}{1} (\bibinfo{year}{2019}).

\bibitem[{\citenamefont{Teoh et~al.}(2020)\citenamefont{Teoh, Drygala, Melko,
  and Islam}}]{teoh2020machine}
\bibinfo{author}{\bibfnamefont{Y.~H.} \bibnamefont{Teoh}},
  \bibinfo{author}{\bibfnamefont{M.}~\bibnamefont{Drygala}},
  \bibinfo{author}{\bibfnamefont{R.~G.} \bibnamefont{Melko}}, \bibnamefont{and}
  \bibinfo{author}{\bibfnamefont{R.}~\bibnamefont{Islam}},
  \bibinfo{journal}{Quantum Science and Technology}
  \textbf{\bibinfo{volume}{5}}, \bibinfo{pages}{024001} (\bibinfo{year}{2020}).

\bibitem[{\citenamefont{Shapira et~al.}(2020)\citenamefont{Shapira, Shaniv,
  Manovitz, Akerman, Peleg, Gazit, Ozeri, and Stern}}]{shapira2020theory}
\bibinfo{author}{\bibfnamefont{Y.}~\bibnamefont{Shapira}},
  \bibinfo{author}{\bibfnamefont{R.}~\bibnamefont{Shaniv}},
  \bibinfo{author}{\bibfnamefont{T.}~\bibnamefont{Manovitz}},
  \bibinfo{author}{\bibfnamefont{N.}~\bibnamefont{Akerman}},
  \bibinfo{author}{\bibfnamefont{L.}~\bibnamefont{Peleg}},
  \bibinfo{author}{\bibfnamefont{L.}~\bibnamefont{Gazit}},
  \bibinfo{author}{\bibfnamefont{R.}~\bibnamefont{Ozeri}}, \bibnamefont{and}
  \bibinfo{author}{\bibfnamefont{A.}~\bibnamefont{Stern}},
  \bibinfo{journal}{Physical Review A} \textbf{\bibinfo{volume}{101}},
  \bibinfo{pages}{032330} (\bibinfo{year}{2020}).

\bibitem[{\citenamefont{Manovitz et~al.}(2020)\citenamefont{Manovitz, Shapira,
  Akerman, Stern, and Ozeri}}]{manovitz2020quantum}
\bibinfo{author}{\bibfnamefont{T.}~\bibnamefont{Manovitz}},
  \bibinfo{author}{\bibfnamefont{Y.}~\bibnamefont{Shapira}},
  \bibinfo{author}{\bibfnamefont{N.}~\bibnamefont{Akerman}},
  \bibinfo{author}{\bibfnamefont{A.}~\bibnamefont{Stern}}, \bibnamefont{and}
  \bibinfo{author}{\bibfnamefont{R.}~\bibnamefont{Ozeri}},
  \bibinfo{journal}{PRX quantum} \textbf{\bibinfo{volume}{1}},
  \bibinfo{pages}{020303} (\bibinfo{year}{2020}).

\bibitem[{\citenamefont{Shapira et~al.}(2023)\citenamefont{Shapira, Manovitz,
  Akerman, Stern, and Ozeri}}]{shapira2023quantum}
\bibinfo{author}{\bibfnamefont{Y.}~\bibnamefont{Shapira}},
  \bibinfo{author}{\bibfnamefont{T.}~\bibnamefont{Manovitz}},
  \bibinfo{author}{\bibfnamefont{N.}~\bibnamefont{Akerman}},
  \bibinfo{author}{\bibfnamefont{A.}~\bibnamefont{Stern}}, \bibnamefont{and}
  \bibinfo{author}{\bibfnamefont{R.}~\bibnamefont{Ozeri}},
  \bibinfo{journal}{Physical Review X} \textbf{\bibinfo{volume}{13}},
  \bibinfo{pages}{021021} (\bibinfo{year}{2023}).

\bibitem[{\citenamefont{Scholl et~al.}(2022)\citenamefont{Scholl, Williams,
  Bornet, Wallner, Barredo, Henriet, Signoles, Hainaut, Franz, Geier
  et~al.}}]{scholl2022microwave}
\bibinfo{author}{\bibfnamefont{P.}~\bibnamefont{Scholl}},
  \bibinfo{author}{\bibfnamefont{H.~J.} \bibnamefont{Williams}},
  \bibinfo{author}{\bibfnamefont{G.}~\bibnamefont{Bornet}},
  \bibinfo{author}{\bibfnamefont{F.}~\bibnamefont{Wallner}},
  \bibinfo{author}{\bibfnamefont{D.}~\bibnamefont{Barredo}},
  \bibinfo{author}{\bibfnamefont{L.}~\bibnamefont{Henriet}},
  \bibinfo{author}{\bibfnamefont{A.}~\bibnamefont{Signoles}},
  \bibinfo{author}{\bibfnamefont{C.}~\bibnamefont{Hainaut}},
  \bibinfo{author}{\bibfnamefont{T.}~\bibnamefont{Franz}},
  \bibinfo{author}{\bibfnamefont{S.}~\bibnamefont{Geier}},
  \bibnamefont{et~al.}, \bibinfo{journal}{PRX Quantum}
  \textbf{\bibinfo{volume}{3}}, \bibinfo{pages}{020303} (\bibinfo{year}{2022}).

\bibitem[{\citenamefont{Debnath et~al.}(2016)\citenamefont{Debnath, Linke,
  Figgatt, Landsman, Wright, and Monroe}}]{Debnath2016}
\bibinfo{author}{\bibfnamefont{S.}~\bibnamefont{Debnath}},
  \bibinfo{author}{\bibfnamefont{N.~M.} \bibnamefont{Linke}},
  \bibinfo{author}{\bibfnamefont{C.}~\bibnamefont{Figgatt}},
  \bibinfo{author}{\bibfnamefont{K.~A.} \bibnamefont{Landsman}},
  \bibinfo{author}{\bibfnamefont{K.}~\bibnamefont{Wright}}, \bibnamefont{and}
  \bibinfo{author}{\bibfnamefont{C.}~\bibnamefont{Monroe}},
  \bibinfo{journal}{Nature} \textbf{\bibinfo{volume}{536}}, \bibinfo{pages}{63}
  (\bibinfo{year}{2016}), ISSN \bibinfo{issn}{0028-0836}.

\bibitem[{\citenamefont{Pino et~al.}(2021)\citenamefont{Pino, Dreiling,
  Figgatt, Gaebler, Moses, Allman, Baldwin, Foss-Feig, Hayes, Mayer
  et~al.}}]{Honeywell_QCCD}
\bibinfo{author}{\bibfnamefont{J.~M.} \bibnamefont{Pino}},
  \bibinfo{author}{\bibfnamefont{J.~M.} \bibnamefont{Dreiling}},
  \bibinfo{author}{\bibfnamefont{C.}~\bibnamefont{Figgatt}},
  \bibinfo{author}{\bibfnamefont{J.~P.} \bibnamefont{Gaebler}},
  \bibinfo{author}{\bibfnamefont{S.~A.} \bibnamefont{Moses}},
  \bibinfo{author}{\bibfnamefont{M.}~\bibnamefont{Allman}},
  \bibinfo{author}{\bibfnamefont{C.}~\bibnamefont{Baldwin}},
  \bibinfo{author}{\bibfnamefont{M.}~\bibnamefont{Foss-Feig}},
  \bibinfo{author}{\bibfnamefont{D.}~\bibnamefont{Hayes}},
  \bibinfo{author}{\bibfnamefont{K.}~\bibnamefont{Mayer}},
  \bibnamefont{et~al.}, \bibinfo{journal}{Nature}
  \textbf{\bibinfo{volume}{592}}, \bibinfo{pages}{209} (\bibinfo{year}{2021}).

\bibitem[{\citenamefont{Martinez et~al.}(2016)\citenamefont{Martinez, Muschik,
  Schindler, Nigg, Erhard, Heyl, Hauke, Dalmonte, Monz, Zoller
  et~al.}}]{martinez2016real}
\bibinfo{author}{\bibfnamefont{E.~A.} \bibnamefont{Martinez}},
  \bibinfo{author}{\bibfnamefont{C.~A.} \bibnamefont{Muschik}},
  \bibinfo{author}{\bibfnamefont{P.}~\bibnamefont{Schindler}},
  \bibinfo{author}{\bibfnamefont{D.}~\bibnamefont{Nigg}},
  \bibinfo{author}{\bibfnamefont{A.}~\bibnamefont{Erhard}},
  \bibinfo{author}{\bibfnamefont{M.}~\bibnamefont{Heyl}},
  \bibinfo{author}{\bibfnamefont{P.}~\bibnamefont{Hauke}},
  \bibinfo{author}{\bibfnamefont{M.}~\bibnamefont{Dalmonte}},
  \bibinfo{author}{\bibfnamefont{T.}~\bibnamefont{Monz}},
  \bibinfo{author}{\bibfnamefont{P.}~\bibnamefont{Zoller}},
  \bibnamefont{et~al.}, \bibinfo{journal}{Nature}
  \textbf{\bibinfo{volume}{534}}, \bibinfo{pages}{516} (\bibinfo{year}{2016}).

\bibitem[{wu2()}]{wu2022Hamiltoniansupplemental}
\bibinfo{note}{See supplemental materials.}

\bibitem[{\citenamefont{Wu et~al.}(2023)\citenamefont{Wu, Shi, and
  Zhang}}]{wu2023continuous}
\bibinfo{author}{\bibfnamefont{Q.}~\bibnamefont{Wu}},
  \bibinfo{author}{\bibfnamefont{Y.}~\bibnamefont{Shi}}, \bibnamefont{and}
  \bibinfo{author}{\bibfnamefont{J.}~\bibnamefont{Zhang}},
  \bibinfo{journal}{Physical Review Research} \textbf{\bibinfo{volume}{5}},
  \bibinfo{pages}{023022} (\bibinfo{year}{2023}).

\bibitem[{\citenamefont{Wu}(2023)}]{wu2023quantum}
\bibinfo{author}{\bibfnamefont{Q.}~\bibnamefont{Wu}}, Ph.D. thesis,
  \bibinfo{school}{New York University} (\bibinfo{year}{2023}).

\bibitem[{\citenamefont{Porras and Cirac}(2004)}]{porras2004effective}
\bibinfo{author}{\bibfnamefont{D.}~\bibnamefont{Porras}} \bibnamefont{and}
  \bibinfo{author}{\bibfnamefont{J.~I.} \bibnamefont{Cirac}},
  \bibinfo{journal}{Physical review letters} \textbf{\bibinfo{volume}{92}},
  \bibinfo{pages}{207901} (\bibinfo{year}{2004}).

\bibitem[{\citenamefont{Sch{\"a}fer et~al.}(2018)\citenamefont{Sch{\"a}fer,
  Ballance, Thirumalai, Stephenson, Ballance, Steane, and
  Lucas}}]{schafer2018fast}
\bibinfo{author}{\bibfnamefont{V.}~\bibnamefont{Sch{\"a}fer}},
  \bibinfo{author}{\bibfnamefont{C.}~\bibnamefont{Ballance}},
  \bibinfo{author}{\bibfnamefont{K.}~\bibnamefont{Thirumalai}},
  \bibinfo{author}{\bibfnamefont{L.}~\bibnamefont{Stephenson}},
  \bibinfo{author}{\bibfnamefont{T.}~\bibnamefont{Ballance}},
  \bibinfo{author}{\bibfnamefont{A.}~\bibnamefont{Steane}}, \bibnamefont{and}
  \bibinfo{author}{\bibfnamefont{D.}~\bibnamefont{Lucas}},
  \bibinfo{journal}{Nature} \textbf{\bibinfo{volume}{555}}, \bibinfo{pages}{75}
  (\bibinfo{year}{2018}).

\bibitem[{\citenamefont{Arrazola et~al.}(2016)\citenamefont{Arrazola,
  Pedernales, Lamata, and Solano}}]{arrazola2016digital}
\bibinfo{author}{\bibfnamefont{I.}~\bibnamefont{Arrazola}},
  \bibinfo{author}{\bibfnamefont{J.~S.} \bibnamefont{Pedernales}},
  \bibinfo{author}{\bibfnamefont{L.}~\bibnamefont{Lamata}}, \bibnamefont{and}
  \bibinfo{author}{\bibfnamefont{E.}~\bibnamefont{Solano}},
  \bibinfo{journal}{Scientific reports} \textbf{\bibinfo{volume}{6}},
  \bibinfo{pages}{1} (\bibinfo{year}{2016}).

\bibitem[{\citenamefont{Olsacher et~al.}(2020)\citenamefont{Olsacher, Postler,
  Schindler, Monz, Zoller, and Sieberer}}]{olsacher2020scalable}
\bibinfo{author}{\bibfnamefont{T.}~\bibnamefont{Olsacher}},
  \bibinfo{author}{\bibfnamefont{L.}~\bibnamefont{Postler}},
  \bibinfo{author}{\bibfnamefont{P.}~\bibnamefont{Schindler}},
  \bibinfo{author}{\bibfnamefont{T.}~\bibnamefont{Monz}},
  \bibinfo{author}{\bibfnamefont{P.}~\bibnamefont{Zoller}}, \bibnamefont{and}
  \bibinfo{author}{\bibfnamefont{L.~M.} \bibnamefont{Sieberer}},
  \bibinfo{journal}{PRX Quantum} \textbf{\bibinfo{volume}{1}},
  \bibinfo{pages}{020316} (\bibinfo{year}{2020}).

\bibitem[{\citenamefont{Ebadi et~al.}(2022)\citenamefont{Ebadi, Keesling, Cain,
  Wang, Levine, Bluvstein, Semeghini, Omran, Liu, Samajdar
  et~al.}}]{ebadi2022quantum}
\bibinfo{author}{\bibfnamefont{S.}~\bibnamefont{Ebadi}},
  \bibinfo{author}{\bibfnamefont{A.}~\bibnamefont{Keesling}},
  \bibinfo{author}{\bibfnamefont{M.}~\bibnamefont{Cain}},
  \bibinfo{author}{\bibfnamefont{T.~T.} \bibnamefont{Wang}},
  \bibinfo{author}{\bibfnamefont{H.}~\bibnamefont{Levine}},
  \bibinfo{author}{\bibfnamefont{D.}~\bibnamefont{Bluvstein}},
  \bibinfo{author}{\bibfnamefont{G.}~\bibnamefont{Semeghini}},
  \bibinfo{author}{\bibfnamefont{A.}~\bibnamefont{Omran}},
  \bibinfo{author}{\bibfnamefont{J.-G.} \bibnamefont{Liu}},
  \bibinfo{author}{\bibfnamefont{R.}~\bibnamefont{Samajdar}},
  \bibnamefont{et~al.}, \bibinfo{journal}{Science}  (\bibinfo{year}{2022}).

\bibitem[{\citenamefont{Gopalakrishnan
  et~al.}(2011)\citenamefont{Gopalakrishnan, Lev, and
  Goldbart}}]{gopalakrishnan2011frustration}
\bibinfo{author}{\bibfnamefont{S.}~\bibnamefont{Gopalakrishnan}},
  \bibinfo{author}{\bibfnamefont{B.~L.} \bibnamefont{Lev}}, \bibnamefont{and}
  \bibinfo{author}{\bibfnamefont{P.~M.} \bibnamefont{Goldbart}},
  \bibinfo{journal}{Physical review letters} \textbf{\bibinfo{volume}{107}},
  \bibinfo{pages}{277201} (\bibinfo{year}{2011}).

\bibitem[{\citenamefont{Korenblit et~al.}(2012)\citenamefont{Korenblit, Kafri,
  Campbell, Islam, Edwards, Gong, Lin, Duan, Kim, Kim
  et~al.}}]{korenblit2012quantum}
\bibinfo{author}{\bibfnamefont{S.}~\bibnamefont{Korenblit}},
  \bibinfo{author}{\bibfnamefont{D.}~\bibnamefont{Kafri}},
  \bibinfo{author}{\bibfnamefont{W.~C.} \bibnamefont{Campbell}},
  \bibinfo{author}{\bibfnamefont{R.}~\bibnamefont{Islam}},
  \bibinfo{author}{\bibfnamefont{E.~E.} \bibnamefont{Edwards}},
  \bibinfo{author}{\bibfnamefont{Z.-X.} \bibnamefont{Gong}},
  \bibinfo{author}{\bibfnamefont{G.-D.} \bibnamefont{Lin}},
  \bibinfo{author}{\bibfnamefont{L.-M.} \bibnamefont{Duan}},
  \bibinfo{author}{\bibfnamefont{J.}~\bibnamefont{Kim}},
  \bibinfo{author}{\bibfnamefont{K.}~\bibnamefont{Kim}}, \bibnamefont{et~al.},
  \bibinfo{journal}{New Journal of Physics} \textbf{\bibinfo{volume}{14}},
  \bibinfo{pages}{095024} (\bibinfo{year}{2012}).

\bibitem[{\citenamefont{Ryan-Anderson et~al.}(2022)\citenamefont{Ryan-Anderson,
  Brown, Allman, Arkin, Asa-Attuah, Baldwin, Berg, Bohnet, Braxton, Burdick
  et~al.}}]{Honeywell_QEC}
\bibinfo{author}{\bibfnamefont{C.}~\bibnamefont{Ryan-Anderson}},
  \bibinfo{author}{\bibfnamefont{N.}~\bibnamefont{Brown}},
  \bibinfo{author}{\bibfnamefont{M.}~\bibnamefont{Allman}},
  \bibinfo{author}{\bibfnamefont{B.}~\bibnamefont{Arkin}},
  \bibinfo{author}{\bibfnamefont{G.}~\bibnamefont{Asa-Attuah}},
  \bibinfo{author}{\bibfnamefont{C.}~\bibnamefont{Baldwin}},
  \bibinfo{author}{\bibfnamefont{J.}~\bibnamefont{Berg}},
  \bibinfo{author}{\bibfnamefont{J.}~\bibnamefont{Bohnet}},
  \bibinfo{author}{\bibfnamefont{S.}~\bibnamefont{Braxton}},
  \bibinfo{author}{\bibfnamefont{N.}~\bibnamefont{Burdick}},
  \bibnamefont{et~al.}, \bibinfo{journal}{arXiv preprint arXiv:2208.01863}
  (\bibinfo{year}{2022}).

\bibitem[{\citenamefont{Kitaev}(2003)}]{kitaev2003fault}
\bibinfo{author}{\bibfnamefont{A.~Y.} \bibnamefont{Kitaev}},
  \bibinfo{journal}{Annals of physics} \textbf{\bibinfo{volume}{303}},
  \bibinfo{pages}{2} (\bibinfo{year}{2003}).

\end{thebibliography}


\begin{thebibliography}{12}
\expandafter\ifx\csname natexlab\endcsname\relax\def\natexlab#1{#1}\fi
\expandafter\ifx\csname bibnamefont\endcsname\relax
  \def\bibnamefont#1{#1}\fi
\expandafter\ifx\csname bibfnamefont\endcsname\relax
  \def\bibfnamefont#1{#1}\fi
\expandafter\ifx\csname citenamefont\endcsname\relax
  \def\citenamefont#1{#1}\fi
\expandafter\ifx\csname url\endcsname\relax
  \def\url#1{\texttt{#1}}\fi
\expandafter\ifx\csname urlprefix\endcsname\relax\def\urlprefix{URL }\fi
\providecommand{\bibinfo}[2]{#2}
\providecommand{\eprint}[2][]{\url{#2}}

\bibitem[{\citenamefont{Shapira et~al.}(2020)\citenamefont{Shapira, Shaniv,
  Manovitz, Akerman, Peleg, Gazit, Ozeri, and Stern}}]{shapira2020theory}
\bibinfo{author}{\bibfnamefont{Y.}~\bibnamefont{Shapira}},
  \bibinfo{author}{\bibfnamefont{R.}~\bibnamefont{Shaniv}},
  \bibinfo{author}{\bibfnamefont{T.}~\bibnamefont{Manovitz}},
  \bibinfo{author}{\bibfnamefont{N.}~\bibnamefont{Akerman}},
  \bibinfo{author}{\bibfnamefont{L.}~\bibnamefont{Peleg}},
  \bibinfo{author}{\bibfnamefont{L.}~\bibnamefont{Gazit}},
  \bibinfo{author}{\bibfnamefont{R.}~\bibnamefont{Ozeri}}, \bibnamefont{and}
  \bibinfo{author}{\bibfnamefont{A.}~\bibnamefont{Stern}},
  \bibinfo{journal}{Physical Review A} \textbf{\bibinfo{volume}{101}},
  \bibinfo{pages}{032330} (\bibinfo{year}{2020}).

\bibitem[{\citenamefont{Swingle}(2012)}]{swingle2012entanglement}
\bibinfo{author}{\bibfnamefont{B.}~\bibnamefont{Swingle}},
  \bibinfo{journal}{Physical Review D} \textbf{\bibinfo{volume}{86}},
  \bibinfo{pages}{065007} (\bibinfo{year}{2012}).

\bibitem[{\citenamefont{Swingle}(2018)}]{swingle2018spacetime}
\bibinfo{author}{\bibfnamefont{B.}~\bibnamefont{Swingle}},
  \bibinfo{journal}{Annual Review of Condensed Matter Physics}
  \textbf{\bibinfo{volume}{9}}, \bibinfo{pages}{345} (\bibinfo{year}{2018}).

\bibitem[{\citenamefont{Hayden and Preskill}(2007)}]{hayden2007black}
\bibinfo{author}{\bibfnamefont{P.}~\bibnamefont{Hayden}} \bibnamefont{and}
  \bibinfo{author}{\bibfnamefont{J.}~\bibnamefont{Preskill}},
  \bibinfo{journal}{Journal of high energy physics}
  \textbf{\bibinfo{volume}{2007}}, \bibinfo{pages}{120} (\bibinfo{year}{2007}).

\bibitem[{\citenamefont{Sekino and Susskind}(2008)}]{sekino2008fast}
\bibinfo{author}{\bibfnamefont{Y.}~\bibnamefont{Sekino}} \bibnamefont{and}
  \bibinfo{author}{\bibfnamefont{L.}~\bibnamefont{Susskind}},
  \bibinfo{journal}{Journal of High Energy Physics}
  \textbf{\bibinfo{volume}{2008}}, \bibinfo{pages}{065} (\bibinfo{year}{2008}).

\bibitem[{\citenamefont{Kitaev}(2003)}]{kitaev2003fault}
\bibinfo{author}{\bibfnamefont{A.~Y.} \bibnamefont{Kitaev}},
  \bibinfo{journal}{Annals of physics} \textbf{\bibinfo{volume}{303}},
  \bibinfo{pages}{2} (\bibinfo{year}{2003}).

\bibitem[{\citenamefont{Lhuillier and
  Misguich}(2010)}]{lhuillier2010introduction}
\bibinfo{author}{\bibfnamefont{C.}~\bibnamefont{Lhuillier}} \bibnamefont{and}
  \bibinfo{author}{\bibfnamefont{G.}~\bibnamefont{Misguich}}, in
  \emph{\bibinfo{booktitle}{Introduction to frustrated magnetism: materials,
  experiments, theory}} (\bibinfo{publisher}{Springer}, \bibinfo{year}{2010}),
  pp. \bibinfo{pages}{23--41}.

\bibitem[{\citenamefont{Eggarter}(1974)}]{eggarter1974cayley}
\bibinfo{author}{\bibfnamefont{T.}~\bibnamefont{Eggarter}},
  \bibinfo{journal}{Physical Review B} \textbf{\bibinfo{volume}{9}},
  \bibinfo{pages}{2989} (\bibinfo{year}{1974}).

\bibitem[{\citenamefont{Bentsen et~al.}(2019)\citenamefont{Bentsen, Hashizume,
  Buyskikh, Davis, Daley, Gubser, and Schleier-Smith}}]{bentsen2019treelike}
\bibinfo{author}{\bibfnamefont{G.}~\bibnamefont{Bentsen}},
  \bibinfo{author}{\bibfnamefont{T.}~\bibnamefont{Hashizume}},
  \bibinfo{author}{\bibfnamefont{A.~S.} \bibnamefont{Buyskikh}},
  \bibinfo{author}{\bibfnamefont{E.~J.} \bibnamefont{Davis}},
  \bibinfo{author}{\bibfnamefont{A.~J.} \bibnamefont{Daley}},
  \bibinfo{author}{\bibfnamefont{S.~S.} \bibnamefont{Gubser}},
  \bibnamefont{and}
  \bibinfo{author}{\bibfnamefont{M.}~\bibnamefont{Schleier-Smith}},
  \bibinfo{journal}{Physical review letters} \textbf{\bibinfo{volume}{123}},
  \bibinfo{pages}{130601} (\bibinfo{year}{2019}).

\bibitem[{\citenamefont{Baxter}(2016)}]{baxter2016exactly}
\bibinfo{author}{\bibfnamefont{R.~J.} \bibnamefont{Baxter}},
  \emph{\bibinfo{title}{Exactly solved models in statistical mechanics}}
  (\bibinfo{publisher}{Elsevier}, \bibinfo{year}{2016}).

\bibitem[{\citenamefont{Song et~al.}(2021)\citenamefont{Song, Kim, Hwang, Lee,
  and Ahn}}]{Qsim_cayley}
\bibinfo{author}{\bibfnamefont{Y.}~\bibnamefont{Song}},
  \bibinfo{author}{\bibfnamefont{M.}~\bibnamefont{Kim}},
  \bibinfo{author}{\bibfnamefont{H.}~\bibnamefont{Hwang}},
  \bibinfo{author}{\bibfnamefont{W.}~\bibnamefont{Lee}}, \bibnamefont{and}
  \bibinfo{author}{\bibfnamefont{J.}~\bibnamefont{Ahn}},
  \bibinfo{journal}{Phys. Rev. Res.} \textbf{\bibinfo{volume}{3}},
  \bibinfo{pages}{013286} (\bibinfo{year}{2021}).

\bibitem[{\citenamefont{Wu et~al.}(2023)\citenamefont{Wu, Shi, and
  Zhang}}]{wu2023continuous}
\bibinfo{author}{\bibfnamefont{Q.}~\bibnamefont{Wu}},
  \bibinfo{author}{\bibfnamefont{Y.}~\bibnamefont{Shi}}, \bibnamefont{and}
  \bibinfo{author}{\bibfnamefont{J.}~\bibnamefont{Zhang}},
  \bibinfo{journal}{Physical Review Research} \textbf{\bibinfo{volume}{5}},
  \bibinfo{pages}{023022} (\bibinfo{year}{2023}).

\end{thebibliography}

\bibliographystyle{apsrev}

\end{document}


\title{Supplemental material: Qubits on programmable geometries with a trapped-ion quantum processor}

\author{Qiming Wu}
\email{qiming.wu@nyu.edu}
\affiliation{Department of Physics, University of California, Berkeley, CA, USA}
\author{Yue Shi}
\affiliation{Department of Physics, Princeton University, Princeton, NJ, USA}
\author{Jiehang Zhang}
\email{jzhang2022@ustc.edu.cn}
\affiliation{1. School of Physical Sciences, University of Science and Technology of China, Hefei 230026, China
}
\affiliation{2. Shanghai Research Center for Quantum Science and CAS Center for Excellence in Quantum Information and Quantum Physics, University of Science and Technology of China, Shanghai 201315, China}
\affiliation{3. Hefei National Laboratory, University of Science and Technology of China, Hefei 230088, China}
\date{\today}

\renewcommand{\theequation}{S\arabic{equation}}
\renewcommand{\thefigure}{S\arabic{figure}}
\renewcommand{\bibnumfmt}[1]{[S#1]}
\renewcommand{\citenumfont}[1]{S#1}

\maketitle




\title{Supplmentary materials: Qubits on programmable geometries with a trapped-ion quantum processor}


\date{}


\baselineskip24pt


\maketitle 
\section{Designing the target Hamiltonian and fidelity estimation }
\label{sec:fidelity estimation}

In this section, we detail the construction of the target Hamiltonian, and calculate the fidelity to quantify how well the desired Hamiltonians are implemented.  
The target Hamiltonian is generated by summing over a certain set of modes to create the desired interaction. Our method is a generalization of the theory proposal in Ref.~\cite{shapira2020theory}. While numerical optimizations can be used for closing the phonon phase spaces, here we choose a simpler but more scalable method: utilizing the nearly even-spaced axial mode frequencies, driving close to one mode with detunings chosen to be a integer fraction of the mode spacing, we can approximately close the phonon loops simultaneously at the end of each interaction layer. The pulse sequences of high dimensional interaction with 4,6,8 qubits are listed in Table~\ref{tab:Sequence}.

\begin{table}[]

\begin{minipage}[t]{0.45\textwidth} \centering
    4 ions\\
    \vspace{2 mm}
    \begin{tabular}{ |c|c|c| } 
        \hline
        \textbf{Mode} & \textbf{Detuning} & \textbf{Time} \\ 
        \hline
        1 & 107.6 kHz & 9.29 $\mu s$ \\ 
        \hline
        3 & -71.14 kHz & 14.06 $\mu s$\\
        \hline
    \end{tabular}
\end{minipage}
\begin{minipage}[t]{0.45\textwidth} \centering
    6 ions\\
    \vspace{2 mm}
    \begin{tabular}{ |c|c|c| } 
        \hline
        \textbf{Mode} & \textbf{Detuning} & \textbf{Time} \\ 
        \hline
        1 & 107.6 kHz & 9.29 $\mu s$ \\ 
        \hline
        3 & -98.0 kHz & 10.2 $\mu s$\\
        \hline
        5 & -75.6 kHz & 13.2 $\mu s$\\
        \hline
    \end{tabular}
\end{minipage}
\\
\begin{minipage}[t]{\textwidth} \centering
    8 ions\\
    \vspace{2 mm}
    \begin{tabular}{ |c|c|c| } 
        \hline
        \textbf{Mode} & \textbf{Detuning} & \textbf{Time} \\ 
        \hline
        1 & 89.7 kHz & 11.2 $\mu s$ \\ 
        \hline
        3 & -100 kHz & 10.0 $\mu s$\\
        \hline
        5 & -76.7 kHz & 13.0 $\mu s$\\
        \hline
        7 & -70.4 kHz & 14.2 $\mu s$\\
        \hline
   
    \end{tabular}

\end{minipage}
     \caption{Pulses sequences of Hamiltonian engineering with 4,6,8 ions. The detuning and interaction time of each layer is listed. }
    \label{tab:Sequence}
\end{table}

Following the definitions in Ref.~\cite{shapira2020theory},  the fidelity of desired Hamiltonian under the ideal implementation is given by the normalized overlap~\cite{shapira2020theory}:
\begin{equation}
    \mathcal{F} = \frac{1}{2} \left( 1+\frac{\langle J_{ij}^i,J_{ij}^d\rangle}{\sqrt{\langle J_{ij}^i,J_{ij}^i\rangle\langle J_{ij}^d,J_{ij}^d\rangle}}\right),
    \label{eqn:fidelity}
\end{equation}
where $J_{ij}^i$ is the spin-spin interaction strength generated in the experiment under ideal conditions, and $J_{ij}^d$ is the desired interaction strength with non-trivial high-dimensional geometries. 

\begin{figure}
    \centering
    \includegraphics[width = \textwidth]{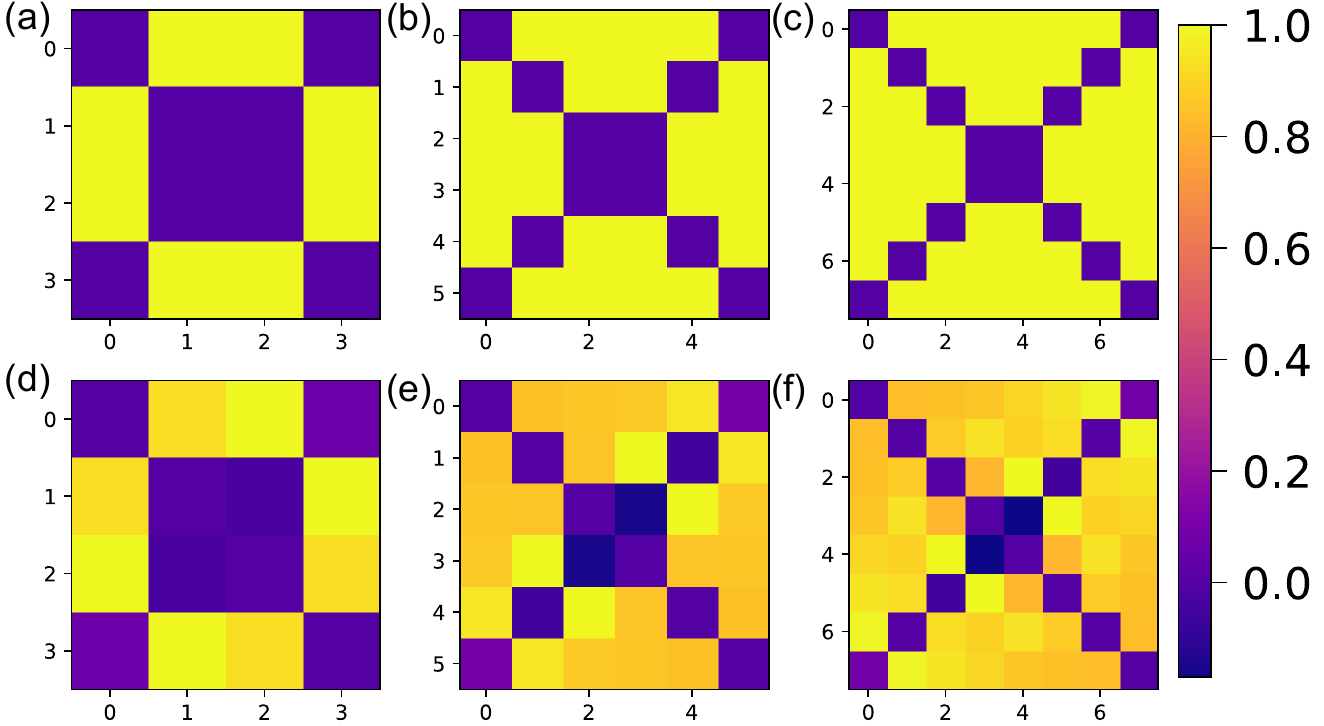}
    \caption{Coupling matrices of high-dimensional interactions are demonstrated in the main text. (a) Desired and (d) experimentally implemented coupling matrix of the 4 qubits on a 2D sphere. The calculated fidelity $\mathcal{F}=99.93\%$. (b) Desired and (e) experimentally implemented coupling matrix of the 6 qubits on a 3D sphere. The calculated fidelity $\mathcal{F}=99.81\%$. (c) Desired and (f) experimentally implemented coupling matrix of the 8 qubits on a 4D sphere. The calculated fidelity $\mathcal{F}=99.86\%$. }
    \label{fig:Jij_graph_468qubit}
\end{figure}

Fig.~\ref{fig:Jij_graph_468qubit} shows the coupling matrices of the desired Hamiltonians and the experimentally implemented Hamiltonians. The maximum interaction strength is normalized to 1 for direct comparisons. Using Eqn.~\ref{eqn:fidelity}, we calculate the Hamiltonian fidelities of the three cases to be $\mathcal{F}=99.93\%, 99.81\%$ and $99.86\%$, respectively.

\section{Qubit on different geometries}
Besides the examples of (hyper-)spheres, a case of positively-curved elliptical geometry presented in the main text, we discuss two other kinds of interactions with non-trivial geometries realizable with our analog-digital hybrid quantum processor and present numerical results. The first type is a tree-like structure as an example of hyperbolic geometry, which is interesting in models of quantum gravity and quantum information~\cite{swingle2012entanglement,swingle2018spacetime,hayden2007black,sekino2008fast}. The second type is a triangular lattice on a torus with 9 qubits, with relevance to quantum error correction~\cite{kitaev2003fault} and topological orders\cite{lhuillier2010introduction}. We numerically found adding a quartic potential on the axial confinement field helps with generating these interactions with high fidelities. 

\subsection{Treelike structures}


Here we calculate ways to generate 1) a Cayley tree~\cite{eggarter1974cayley} with leaves and nodes connected by equal-weight edges and 2) a tunable "leaves-only" tree-like structures for which the points on the real line are best viewed as leaves on an infinite regular tree graphs\cite{bentsen2019treelike}. 

\begin{figure}[!t]
    \centering
    \includegraphics[width = \textwidth]{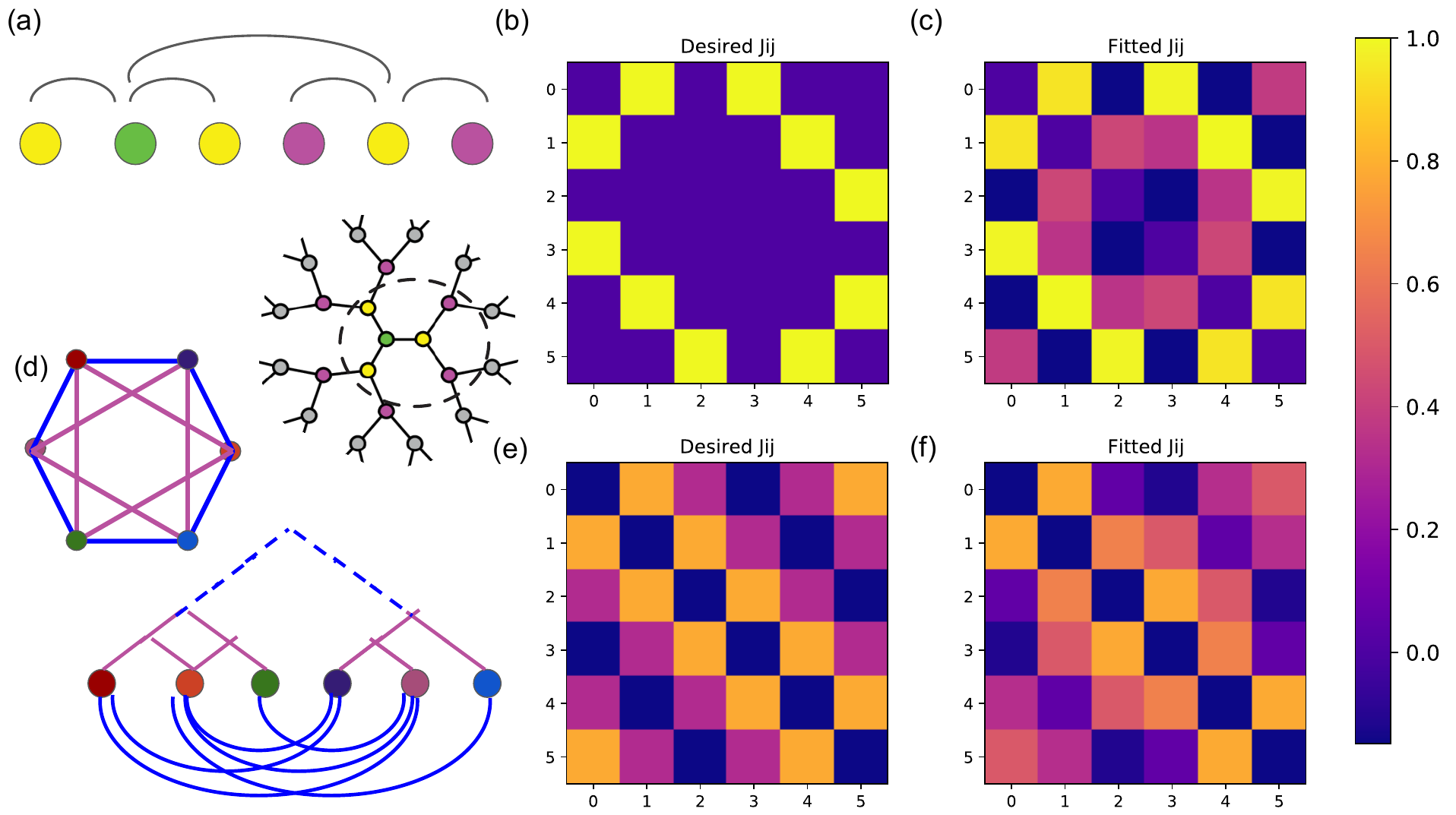}
    \caption{Interaction of the 6 qubits tree-like graphs. Interaction of a 6 qubit. (a) Illustration of interaction on a C3,6 Cayley tree with equal weights. (b) The desired and (c) The calculated $J_{ij}$ matrices of (a) via our Hamiltonian engineering methods. The calculated fidelity is $\mathcal{F}=95.68\%$.  (d) Illustration of tree-like interaction in 2D(See text). The two different coupling strengths are colored blue and purple, respectively. The interaction can be illustrated by a polygon ($s<0$) or a tree with only leaves ($s>0$) and converges to a Bruhat-Tits tree when $s>>1$. The dashed blue lines conclude the blue interactions between two bigger trees. They are further illustrated by the solid blue lines(e) The desired and (f) The calculated $J_{ij}$ matrices of (d) when s= -2 via our Hamiltonian engineering methods. The calculated fidelity $\mathcal{F}=98.69\%$.  }
    \label{fig:tree_graph_combined}
\end{figure}

1): Cayley trees are homogeneous and
isotropic tree graphs in which each non-leaf vertex has a constant number of branches, as illustrated by Fig.~\ref{fig:tree_graph_combined}. To the infinite vertices limit, a Cayley tree is referred to as a Bethe lattice, and it is widely used to describe a unique topology that generates exactly solvable models in classical and quantum systems\cite{baxter2016exactly}. Although the spin dynamics of small size Cayley tree has been studied on Rydberg atoms in tweezer arrays\cite{Qsim_cayley}, here we propose an alternative implementation of a Cayley tree of 6 qubits only using global interactions.

2): The interaction matrix of a "leaves-only" tree-like structure can be written as:
\begin{equation}\label{Eqn:tree_like_Jij}
    J_{ij} = 
    \begin{cases}
    J2^{ls},& \text{if } |i-j|=2^l,l=0,1,2,3...\\
    0             & \text{otherwise,}
\end{cases}
\end{equation}
for which $J$ is the coefficient of the coupling strength, $l$ is an integer number that limits the interaction to occur only at a distance of 2's power and is $s$ a tunable parameter. 
With our mode engineering method, we can tune the value of $s$ in Eqn.~\ref{Eqn:tree_like_Jij} Here we exemplify the graph with $s=-2$ as illustrated by Fig.~\ref{fig:tree_graph_combined} (d) and reach a fidelity of $98.69\%$.

\subsection{Triangular lattice on a torus}
\begin{figure}[!t]
    \centering
    \includegraphics[width = \textwidth]{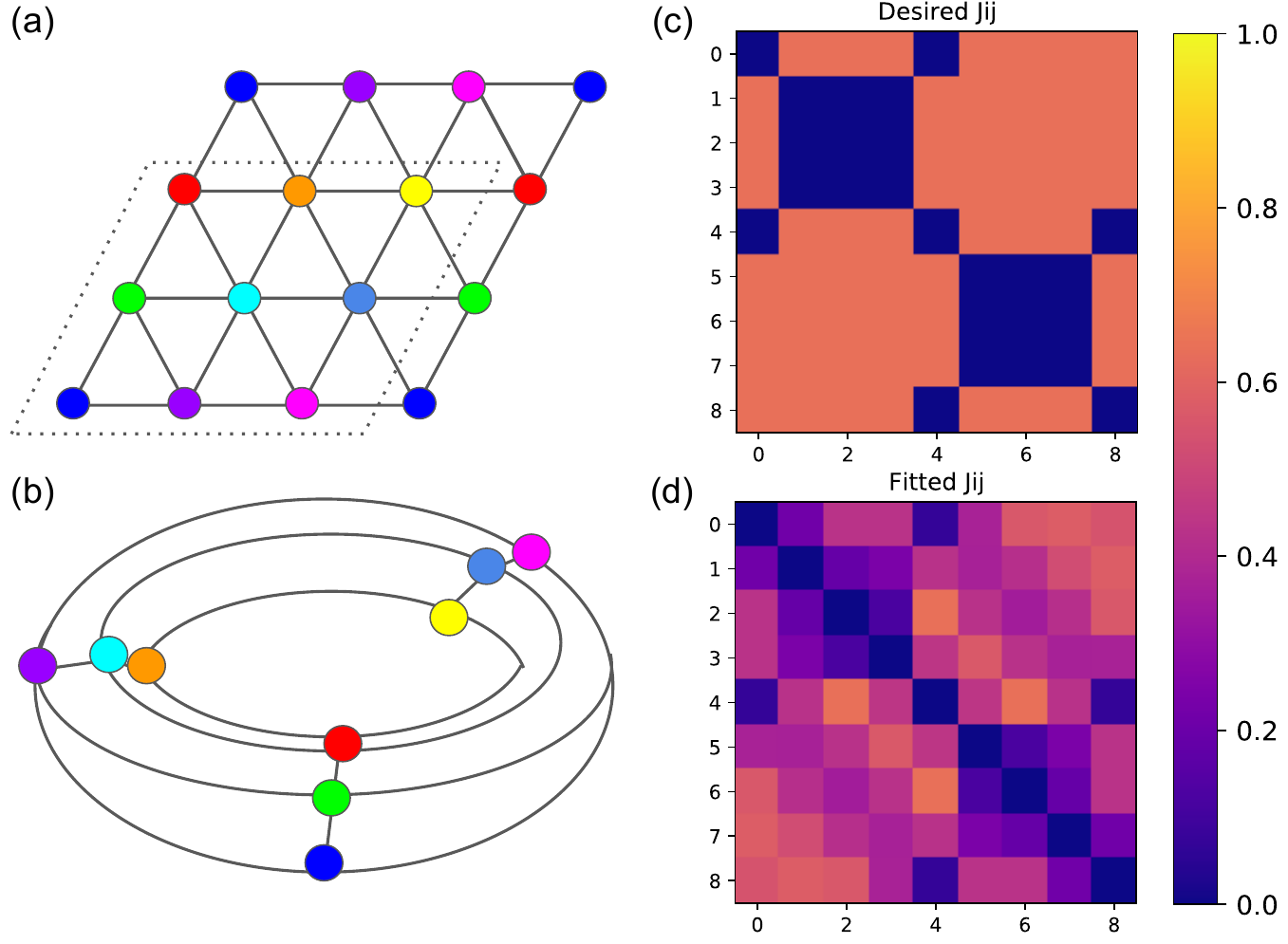}
    \caption{Interaction of a 3 by 3 triangular lattice on a torus. Every color represents a distinct spin. (a) the representation on a 2-dimensional periodic lattice; (b) the representation on a torus Jij Matrix of the equal weight $\sigma_{\phi}\sigma_{\phi}$ of the 9 qubits triangular lattice on a torus. $J_{ij}$ matrices of (c) the desired and (d) the calculated interaction via our Hamiltonian engineering methods. The calculated fidelity is $\mathcal{F}=98.3\%$.  }
    
    \label{fig:9_qubit_triangular_lattice}
\end{figure}

Toric code~\cite{kitaev2003fault} is a topological quantum error correction code on a two-dimensional spin lattice with periodic boundary conditions on both dimensions. Geometrically the lattice forms a ring torus shape, with potential applications in quantum error correction, quantum simulations for different topological orders and spin liquids\cite{kitaev2003fault,lhuillier2010introduction}.  Here we generate an equal weight triangular $\sigma_{\phi}\sigma_{\phi}$ interaction for 9 qubits (3 by 3)  on a triangular lattice as shown by Fig.~\ref{fig:9_qubit_triangular_lattice}: the 9 spins are pairwise entangled according to (a) and encoded on the torus as illustrated by (b). Such interaction is characterized by the interaction matrix shown in (c) a and an experimental realizable Jij matrix with our mode engineering methods shown in (d). We calculate a  fidelity of $98.3\%$ with our approach. 

\section{Experimental calibrations and systematic offsets}
In this section, we discuss the details of experimental calibration and systematic offsets of our trapped ion quantum processor. First, we drive multi-ion blue-sidebands as an important intermediate calibration step for generating the desired interactions. Then, we consider experimental offsets including qubit frequency differences from the magnetic field gradient, and Rabi frequency gradient from the limited Gaussian beam size. We then conclude by numerical simulation to show a comparison of average spin dynamics under the desired Hamiltonian, the ideally implemented Hamiltonian, and the implemented Hamiltonian considered the systematic offset.

\subsection{Blue-sideband calibration}

\begin{figure}[!t]
    \centering
    \includegraphics[width = \textwidth]{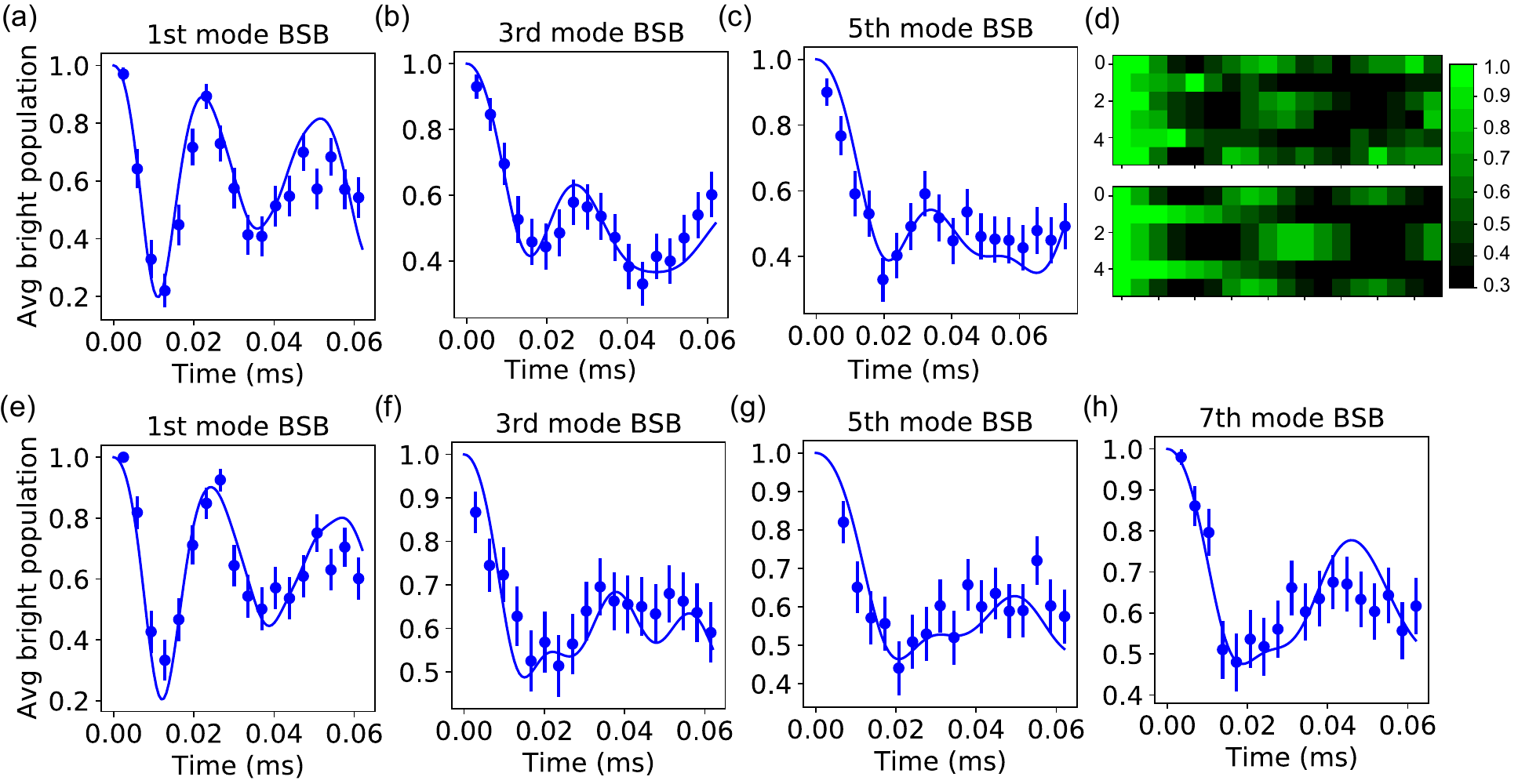}
    \caption{6 and 8 ions BSB floppings on the modes used for Hamiltonian engineering. (a)(b)(c) Measurements of the average population on the $\ket{\downarrow}$ states of BSB time evolution of a 6-ion chain on mode 1,3,5. (d) Experimental (top) and theoretical (bottom) time evolution of individual ion in BSB flopping on mode 3. (e)(f)(g)(h) Measurements of the average population on the $\ket{\downarrow}$ states of BSB time evolution of an 8-ion chain on mode 1,3,5,7.
The error bars denote one standard deviation, and the blue solid lines are numerical simulation
of BSB floppings, assuming $\overline{n} = 0.1$.}
    
    \label{fig:68ion_BSB}
\end{figure}

We calibrate the experiment before driving the spin-spin interaction by probing the global BSB transition on the modes we use for the M-S interaction.
The Hamiltonian of multi spins coupled to one BSB is:
\begin{equation}
\hat{H}_{\text {bsb }}=  \sum_i \eta_{i, m} \Omega_i  \left( \sigma_{+}^{(i)}{\hat{a_m}}^{\dagger}+ \sigma_{-}^{(i)}\hat{a_m} \right),
\end{equation}
where $\eta_{i,m}$ is the  LD parameter of the $i$-th ion on the $m$-th mode, $\Omega_i$ is the Rabi frequency of the $i$-th ion, $\sigma_{+}^{(i)}$ ($\sigma_{-}^{(i)}$) is the spin-flip operator of the $i$-th ion, $\hat{a_m}$ ($\hat{a_m}^{\dagger}$) is the annihilation (creation) operator on the $m$-th mode.

Fig.~\ref{fig:68ion_BSB} shows the time evolution of the average BSB spin excitation on mode 1,3,5 with 6 ions and on mode 1,3,5,7 with 8 ions, similar to that of Fig.~2(b) in main text. These data are taken under a constant carrier Rabi frequency $\Omega/2\pi$ = 41~kHz. The solid lines show the numerical simulation of BSB flopping, assuming the ions start from $|i\rangle=\ket{\downarrow \downarrow \ldots \downarrow}\otimes \sum_{n} \frac{\overline{n}{ }^{n}}{\left(\overline{n}+1\right)^{n+1}}\left|n\right\rangle\left\langle n\right|$ state at $t=0$. Because the heating rate of a single ion is at $\dot{\overline{n}} \sim 200$ quanta/s, the heating effect in 100 $\mu s$ can be negligible in a chain of less than 10 ions for all the modes. Fig.~\ref{fig:68ion_BSB} (d) shows the time evolution of the individual spin when doing BSB flopping on mode 3 with 6 ions, illustrating the differential couplings of ions to this mode.

\begin{figure}
    \centering
    \includegraphics[width = 0.7\textwidth]{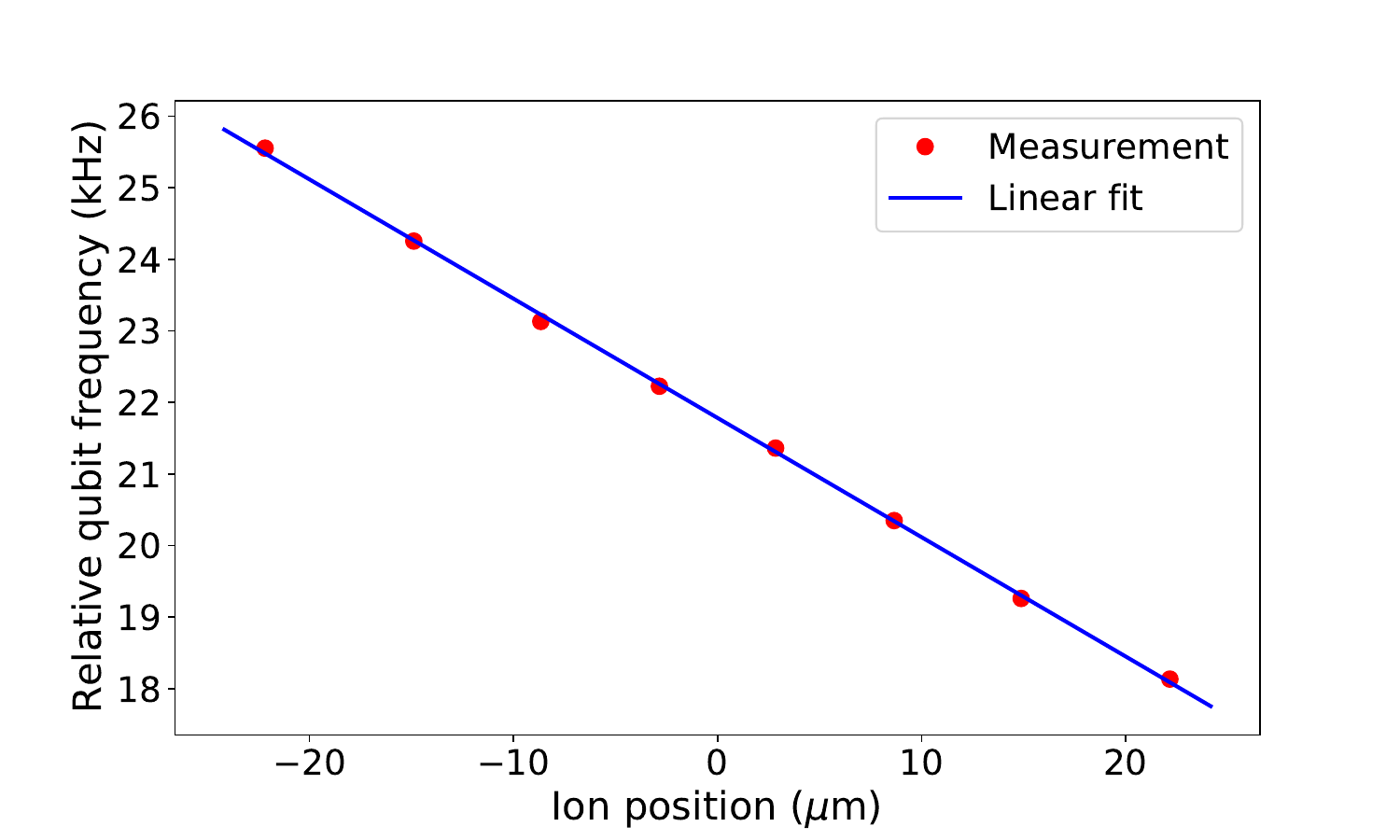}
    \caption{Measurement of qubit frequency gradient using an 8-ion chain. The red dots are measured Ramsey frequencies of individual ions as a function of their relative positions, and the blue solid line is a linear fit.}
    
    \label{fig:qubit_gradient}
\end{figure}

\subsection{Qubit frequency gradient and finite beam size}
Using a microwave field, we perform a global Ramsey experiment of an 8-ion chain. By fitting the Ramsey oscillation frequency of individual ions, we can extract the relative qubit frequency offset as a function of their relative positions.  Fig.~\ref{fig:qubit_gradient} shows the measurement of relative qubit frequencies in an 8-ion chain. The linear fit gives a qubit frequency gradient of 0.167 kHz/$\mathrm{\mu}$m. We consider the qubit frequency gradient as extra $\sigma_z$ terms in addition to the spin-spin interaction in the numerical simulation. The effective Raman laser beam size on the ion plane is measured to be $270\times30$~$\mu$m~\cite{wu2023continuous}. In the numerical simulation, we also consider the difference in Rabi frequency for different ions. 


\subsection{Experimental deviation from the desired Hamiltonian}

\begin{figure}
    \centering
    \includegraphics[width = \textwidth]{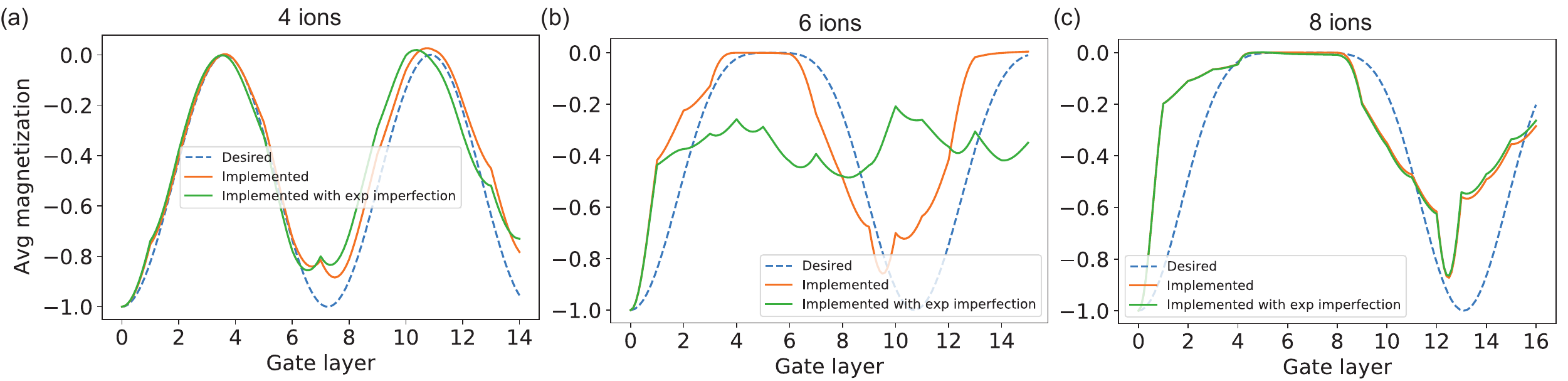}
    \caption{Numerical simulation of the time evolution of the average magnetization of 4,6,8 qubits on high dimensional spaces. Blue dashed, orange solid, green solid denote the dynamics of the desired Hamiltonian, implemented Hamiltonian, and implemented Hamiltonian with systematic imperfection, respectively. }
    
    \label{fig:numerical_comparison}
\end{figure}

We present a numerical comparison of the spin dynamics of the desired Hamiltonian, ideally implemented Hamiltonian, and implemented Hamiltonian taking account of the experimental offsets. The average spin time evolution helps us quantify how well the implementation captures the designed Hamiltonian. Blue dashed lines in Fig.~\ref{fig:numerical_comparison} show the average spin dynamics of the desired Hamiltonian, orange lines show the implemented Hamiltonian as discussed in Section~\ref{sec:fidelity estimation} and green lines show the implemented Hamiltonian with the two inhomogeniety effects. The latter two are simulated using layers of global interaction to reproduce the actual experimental implementation, so the time dynamics are not smooth when interleaving different layers of global interaction. The simulation of the desired Hamiltonians present the time evolution of the total Hamiltonians.   Notably, 
the interaction of 6 ions on a sphere is most susceptible to systematic errors among the three different cases, the exact underlying reasons being not clear at the moment, leaving room for future investigations. 

\bibliography{Hamiltonian_engineering_reference}
\bibliographystyle{apsrev}